%% file: paper.tex
\documentclass[acmsmall]{acmart}
\usepackage[dvipsnames]{xcolor}
\usepackage{url}
\usepackage{pifont}
\usepackage{booktabs}
\usepackage{multirow}
\usepackage{array}
\usepackage{adjustbox}
\usepackage{subcaption}
\usepackage{makecell}
\usepackage{placeins}
\usepackage{xspace}

\usepackage{cleveref} 
\usepackage{mathtools}

\crefname{section}{\S}{\SS}
\crefformat{section}{\S#2#1#3} 
\crefformat{subsection}{\S#2#1#3}
\crefformat{subsubsection}{\S#2#1#3}

\usepackage{enumitem}

\newcommand{\system}{\textsf{ZenFlow}\xspace}
\newcommand{\systemasync}{\textsf{$\system^{*}$}\xspace}

\newcommand{\systemauto}{\textsf{ZenFlow-auto}\xspace}
\newcommand{\systemfull}{\textsf{\system}}

\newcommand{\phead}[1]{\noindent\textbf{#1}} 
\newcommand{\fref}[1]{Fig.~\ref{#1}}
\newcommand{\tref}[1]{Table~\ref{#1}}

\newcommand{\sref}[1]{\cref{#1}}

\begin{document}

\title{ZenFlow: Enabling Stall-Free Offloading for LLM Training}

\author{Tingfeng Lan}
\orcid{0009-0008-5945-6834}
\email{erc8gx@virginia.edu}
\affiliation{%
  \institution{University of Virginia}
  \city{Charlottesville}
  \state{Virginia}
  \country{USA}
}

\author{Yusen Wu}
\orcid{0009-0000-8395-7550}
\email{xrn4ub@virginia.edu}
\affiliation{%
  \institution{University of Virginia}
  \city{Charlottesville}
  \state{Virginia}
  \country{USA}
}

\author{Bin Ma}
\orcid{0009-0003-4287-8946}
\email{bma100@ucmerced.edu}
\affiliation{%
  \institution{University of California, Merced}
  \city{Merced}
  \state{California}
  \country{USA}
}

\author{Zhaoyuan Su}
\orcid{0000-0003-4058-2343}
\email{acf7ea@virginia.edu}
\affiliation{%
  \institution{University of Virginia}
  \city{Charlottesville}
  \state{Virginia}
  \country{USA}
}

\author{Rui Yang}
\orcid{0000-0002-6522-6305}
\email{qgh4hm@virginia.edu}
\affiliation{%
  \institution{University of Virginia}
  \city{Charlottesville}
  \state{Virginia}
  \country{USA}
}

\author{Tekin Bicer}
\orcid{0000-0002-8428-5159}
\email{tbicer@anl.gov}
\affiliation{%
  \institution{Argonne National Laboratory}
  \city{Lemont}
  \state{Illinois}
  \country{USA}
}

\author{Masahiro Tanaka}
\orcid{0009-0001-3972-8133}
\email{tanaka.masahiro@gmail.com}
\affiliation{%
  \institution{AnyScale}
  \city{San Francisco}
  \state{California}
  \country{USA}
}

\author{Olatunji Ruwase}
\orcid{0000-0002-5508-0728}
\email{tunji.ruwase@snowflake.com}
\affiliation{%
  \institution{Snowflake}
  \city{Seattle}
  \state{Washington}
  \country{USA}
}

\author{Dong Li}
\orcid{0000-0001-9336-0694}
\email{dli35@ucmerced.edu}
\affiliation{%
  \institution{University of California, Merced}
  \city{Merced}
  \state{California}
  \country{USA}
}

\author{Yue Cheng}
\orcid{0000-0003-1695-4864}
\email{mrz7dp@virginia.edu}
\affiliation{%
  \institution{University of Virginia}
  \city{Charlottesville}
  \state{Virginia}
  \country{USA}
}

\renewcommand{\shortauthors}{Tingfeng Lan et al.}

\input{contents/abstract}



\setcopyright{cc}
\setcctype{by}
\acmJournal{PACMMOD}
\acmYear{2026} \acmVolume{4} \acmNumber{4 (SIGMOD)} \acmArticle{292}
\acmMonth{9} \acmDOI{10.1145/3837130}

\received{17 January 2026}
\received[revised]{19 May 2026}
\received[accepted]{12 June 2026}

\maketitle

\input{contents/introduction}
\input{contents/background}
\input{contents/overview}
\input{contents/accuracy}
\input{contents/implementation}

\input{contents/evaluation}
\input{contents/related_work}
\input{contents/conclusion}

\begin{acks}
  This work was supported in part by the U.S. National Science Foundation under
  Grants OAC-2411009, CNS-2322860, OAC-2403313, and CCF-2316202. We also acknowledge NSF CloudBank, Modal, and Lambda for providing cloud GPU credits, and the Commonwealth Cyber Initiative (CCI), an
  investment in the advancement of cyber research, innovation, and workforce development. For
  more information about CCI, visit \href{https://cyberinitiative.org}{cyberinitiative.org}.  
\end{acks}

\bibliographystyle{ACM-Reference-Format}
\bibliography{cite}


\end{document}

%% file: contents/abstract.tex
\begin{abstract}

Fine-tuning large language models (LLMs) often exceeds GPU memory limits, prompting systems to offload model states to CPU memory. However, existing offloaded training frameworks like ZeRO-Offload treat all parameters equally and update the full model on the CPU, causing severe GPU stalls, where fast, expensive GPUs sit idle, waiting for slow CPU updates and limited-bandwidth PCIe transfers.

We present {\system}, a new offloaded training framework that decouples updates between GPU and CPU by prioritizing important gradients. {\system} performs in-place updates of important gradients on GPU while offloading and accumulating less important ones on CPU without blocking training iterations, fully overlapping CPU work with GPU computation. To make this importance-aware design scalable, {\system} leverages a novel spatio-temporal locality of important gradients---concentrated in specific positions that remain stable across iterations---to make important gradient identification efficient and lightweight while preserving fidelity. 
Evaluation shows that {\system} achieves up to 5$\times$ end-to-end speedup, 2$\times$ lower PCIe traffic, 85\% less GPU stalls, and up to 72\% lower cloud GPU cost, all while preserving accuracy.

\end{abstract}

%% file: contents/introduction.tex
\section{Introduction}
\label{sec:intro}

Large language models (LLMs) have become the foundation of modern natural language processing applications, powering tasks from text generation to code synthesis~\cite{gpt2_2019,codet5_emnlp2021,chatbot_2020,questionanswering_emnlp2020}. While pretrained models~\cite{gpt2_2019, gpt3_nips20, llama_2023, llama2_2023} offer general-purpose capabilities, fine-tuning them on domain-specific data is often essential for achieving high performance on downstream tasks. 
However, as LLM sizes grow, 
the memory demands of fine-tuning far exceed the capacity of a single GPU,  making large-scale training increasingly challenging~\cite{risingcost_2025,efficienttrainingsurvey_2023, efficient_llm_survey_arxiv24}.

\begin{figure}[t]
    \centering
    \includegraphics[width=0.75\textwidth]{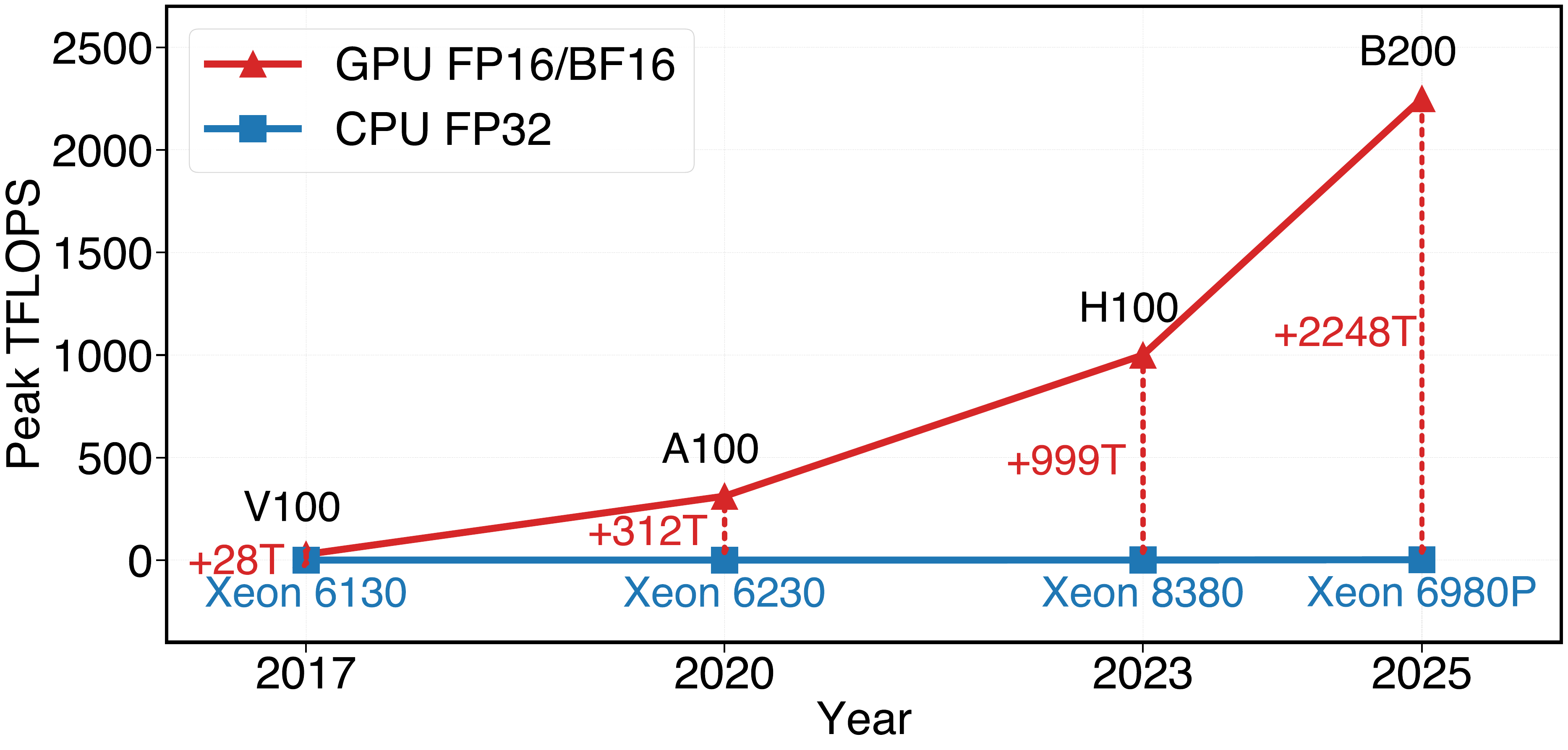} 
    \caption{GPU vs. CPU peak performance evolution. }
    \label{fig:cpu_gpu_perf_gap} 
    \Description{Line chart comparing the growth of peak GPU and CPU floating-point performance from 2016 to 2025.}
\end{figure}

\begin{figure}[ht]
    \centering
    \includegraphics[width=0.75\textwidth]{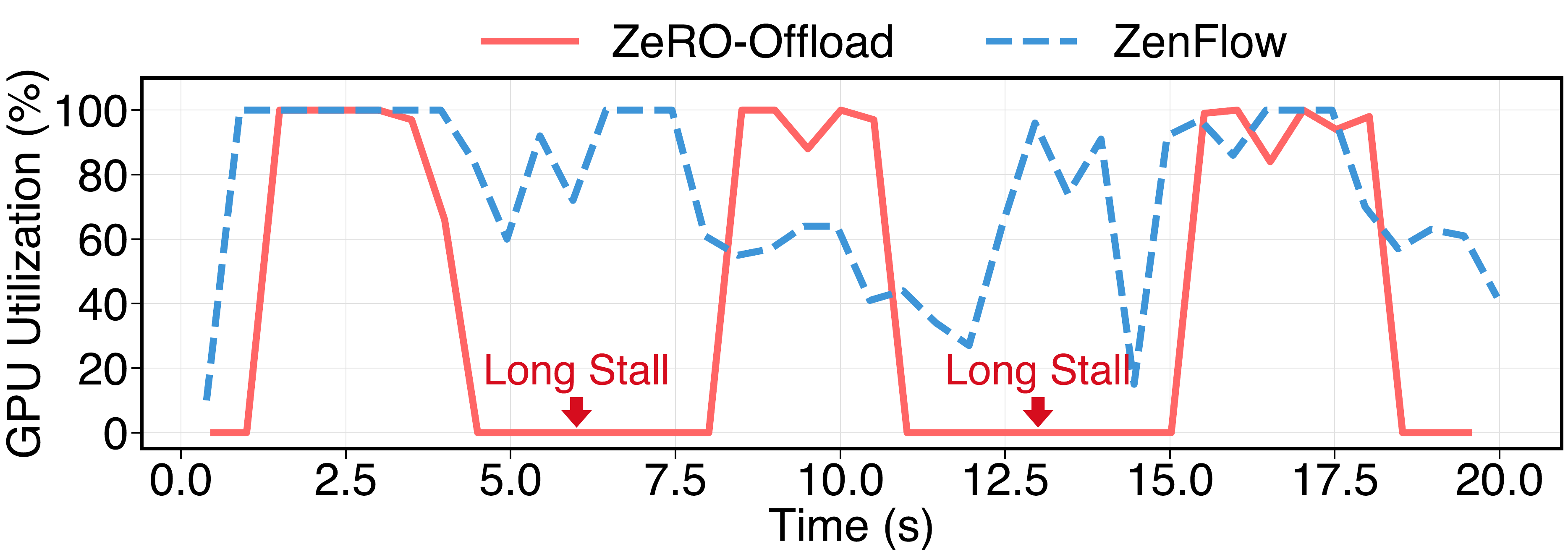}
    \caption{Comparison of GPU utilization over time.}
    \label{fig:gpu_util}
    \Description{GPU utilization timelines for three consecutive training steps under ZeRO-Offload and ZenFlow. ZeRO-Offload repeatedly leaves the GPU idle during CPU-side updates, while ZenFlow keeps GPU utilization high.}

\end{figure}

To address this memory bottleneck, offloading-based training systems, such as ZeRO-Offload~\cite{ren2021zero_offload}, ZeRO-Infinity~\cite{zeroinfinity_sc21}, and StrongHold~\cite{stronghold_sc22}, have emerged as promising solutions. These systems reduce GPU memory consumption by offloading model states (e.g., gradients and optimizer states) to CPU memory or NVMe storage and performing optimizer updates on the CPU. However, the CPU-GPU performance gap has widened rapidly (see \fref{fig:cpu_gpu_perf_gap}), increasing from 28 TFLOPS to 2248 TFLOPS in the last couple of years,
so CPU-side optimizer updates and PCIe transfers are largely exposed to the critical path, causing long GPU stalls (\fref{fig:gpu_util}).
For example, on fine-tuning a Llama-2-7B~\cite{llama2_2023} model with $4\times$ A100 GPUs, one training step time experiences a dramatic increase from 0.5s to 7s when enabling offloading, a 14$\times$ slowdown. 

We identify two major sources of inefficiency in existing offloaded training systems: 
(1)~\emph{Long stalls caused by slow CPU-side updates}, which delay the next training iteration and leave GPUs idle. 
(2)~\emph{High I/O cost}, as each iteration requires transferring the full set of gradients and updated parameters between GPU and CPU.


These performance bottlenecks stem from critical limitations of current offloading systems: 
(1)~They adopt a uniform strategy that treats all parameters and gradients as equally important, regardless of how much each individual gradient actually contributes to learning. 
(2)~They rely on slow CPU with limited computation and PCIe bandwidth for updating and transferring entire model states.

In practice, this causes slow CPU to bottleneck fast and more expensive GPU. As shown in \fref{fig:gpu_util}, during three consecutive training steps, ZeRO-Offload suffers from repeated GPU stalls---each lasting up to 5 seconds within a 7-second step---where GPU utilization drops to nearly 0\%. These prolonged idle periods dominate the training time, resulting in severe underutilization of GPU resources.

However, our analysis and former study~\cite{aji2017sparse,intrinsicdimensionalityexplainseffectiveness_2020,lin2017dgc,jang2024smartinfinityfastlargelanguage} reveal that \emph{LLM gradients are highly imbalanced}: the top 1\% of gradients account for over 90\% of the total gradient norm. Consequently, existing hardware-agnostic treatment of gradients leads to wasted computation and communication on low-impact updates, stalling overall training throughput.  

We introduce \system, the  first offloaded training system 
that decouples parameter updates based on both hardware heterogeneity and learning dynamics (e.g., the gradient importance).  
The key insight behind \system is to \emph{treat important and unimportant updates differently}: by retaining the small but critical subset of gradients on GPU for immediate updates, \system avoids GPU stalls and ensures {\bf fast} updates on important parameters. Meanwhile, remaining less important gradients are offloaded to the {\bf slower} CPU, where they are slowly 
accumulated and updated. This strategy preserves the learning contribution of less important gradients while avoiding frequent update and I/O cost. By amortizing CPU-side updates over multiple GPU iterations, \system effectively combines the high speed of GPU with the cost efficiency of CPU. As illustrated in \fref{fig:gpu_util}, this design enables \system to eliminate repeated GPU stalls seen in ZeRO-Offload, resulting in consistently high GPU utilization and significantly improved end-to-end efficiency. 

A key challenge in realizing \system is determining which parameters are important enough to be updated immediately on GPUs. This becomes particularly difficult in distributed training settings such as ZeRO~\cite{rajbhandari2020zero}, 
where each GPU holds a shard of the full model and its corresponding gradients. Performing an \texttt{AllGather}
to collect the full gradient matrix across GPUs would incur prohibitive communication and memory cost. 
To address this, \system leverages a novel observation: 
\emph{to our best knowledge, we are the first to identify that important gradients in LLM fine-tuning exhibit strong spatial and temporal locality.} 
Specifically, we find that a small subset of input dimensions (i.e., channels) consistently carries high-magnitude gradients across training steps. By tracking and reusing this compact set of important channels, \system enables efficient, scalable importance-aware training without expensive global synchronization.

This paper makes the following contributions:
\begin{itemize}[noitemsep,leftmargin=*]
    \item We discover and characterize a novel spatio-temporal locality property of important gradients in LLM fine-tuning.

    \item We design a lightweight method to identify important gradients in distributed training without costly global synchronization.

    \item We build \system, a fine-grained CPU-GPU pipeline that decouples parameter updates with different importance to minimize GPU stalls and I/O overhead. 

    \item We build \system on DeepSpeed~\cite{deepspeed_kdd20} and evaluate it thoroughly across single-GPU and multi-GPU settings. \system achieves 3.6-5$\times$ end-to-end speedup compared to state-of-the-art offloading systems while maintaining accuracy across diverse LLMs and fine-tuning tasks. 

\end{itemize}

\phead{Open-source release.} 
\emph{{\system} has been integrated into and is open-sourced at 
\href{https://github.com/deepspeedai/DeepSpeed}{\nolinkurl{deepspeedai/DeepSpeed}}.}

%% file: contents/background.tex
\section{Background and Motivation}
\label{sec:background}

\subsection{Distributed Training Systems}
\label{subsec:distributed_training}

\phead{Basics of Distributed Training.} 
Deep learning model training typically consists of millions of iterations performed across multiple training epochs~\cite{gpt3_nips20,llama_2023, llama2_2023}. Each iteration mainly involves three stages: \textit{}forward propagation (\texttt{FP}), backward propagation (\texttt{BP}), and parameter update (\texttt{UP}). In the \texttt{FP} stage, a batch of training data is passed through the model to compute the output and loss based on an objective function.  In the \texttt{BP} stage, the model propagates the loss value reversely through model layers to compute gradients for each model parameter.  Finally, in the \texttt{UP} stage, model parameters are updated using the computed gradients through an optimization algorithm by the optimizer (e.g., SGD~\cite{sgd_2010}, Adam~\cite{adam_2017}).

\phead{Distributed Training.} 
To train large models efficiently, distributed parallelism is widely adopted. When models fit in GPU memory, data parallelism~\cite{ps_osdi2014} is commonly used, replicating the model across devices and sharding/distributing input batches. For larger models that exceed memory capacity, model parallelism~\cite{shoeybi2019megatron} and pipeline parallelism~\cite{gpipe_nips2019, pipedream_sosp2019} partition model layers across devices to utilize aggregate memory. 
Furthermore, ZeRO~\cite{deepspeed_kdd20} extends data parallelism by sharding model parameters, gradients, and optimizer states across devices. ZeRO can be regarded as a hybrid of data and model parallelism: each device processes a full batch in data parallelism style while applying model-parallelism-like partitioning to memory-intensive states. This design enables training models far beyond single-device memory limits and has become foundational in modern systems such as DeepSpeed~\cite{deepspeed_kdd20}, Megatron-LM~\cite{shoeybi2019megatron}, and PyTorch FSDP~\cite{zhao2023pytorch_fsdp}.

\begin{figure*}[t]
    \centering
    \includegraphics[width=\textwidth]{figures/camera_ready/zenflow.pdf}
    \caption{Offloading strategy comparison. (a)~\textbf{ZeRO-Offload}~\cite{ren2021zero_offload} sequentially executes \texttt{FP} and \texttt{BP} on GPU, then offloads gradients and performs \texttt{UP} on CPU, leaving the GPU idle. 
    (b)~\textbf{StrongHold}~\cite{stronghold_sc22} 
    overlaps CPU updates with GPU backward computation and gradient offloading by offloading and updating gradients layer by layer. However, CPU-side update is still too slow to hide, causing GPU stalls. 
    (c)~\textbf{Example: iteration with accumulation:} Update occurs every two steps---one for accumulation (i.e., gradients are summed without applying updates) (\texttt{ACC}), one for normal update (\texttt{UP}). 
    (d)~\textbf{\system} prioritizes parameter updates for important gradient (\sref{subsec:gradient_selection}) on fast GPU to leverage its high compute bandwidth, while accumulating unimportant gradients in several rounds on slow CPU to reduce unnecessary parameter update overhead (\sref{subsec:async_offloading}). This design decouples fast GPU computing from slow CPU processing, minimizing GPU stalls and reducing data movement 
    (\sref{subsec:zero_bubble_pipeline}).} 
    \Description{A chart comparing different offloading methods.}
    \label{fig:offload_comparision}
\end{figure*}

\begin{table}[ht]
    \centering
    \caption{Comparison of fine-tuning approaches.}
    \label{tab:method_comparison}
    \footnotesize
    \begin{adjustbox}{max width=0.90\columnwidth}
    \begin{tabular}{@{}lccc@{}}
    \toprule
    \textbf{Method} & \textbf{Parameter Updates} & \textbf{Memory Cost} & \textbf{Training Speed} \\ \midrule
    Full Fine-tuning 
    & \textcolor{ForestGreen}{Full ($100\%$)} 
    & \textcolor{BrickRed}{Very High (4M)} 
    & \textcolor{BrickRed}{Low} \\
    PEFT (e.g., LoRA)
    & \textcolor{BrickRed}{Partial ($<1\%$)} 
    & \textcolor{ForestGreen}{Low (M)} 
    & \textcolor{ForestGreen}{High} \\
    ZeRO-Offload
    & \textcolor{ForestGreen}{Full ($100\%$)} 
    & \textcolor{ForestGreen}{Low (M)} 
    & \textcolor{BrickRed}{Low} \\
    ZenFlow (ours) 
    & \textcolor{ForestGreen}{Full ($100\%$)} 
    & \textcolor{ForestGreen}{Low (M)}
    & \textcolor{ForestGreen}{High} \\ 
    \bottomrule
    \end{tabular}
    \end{adjustbox}
\end{table}

\begin{figure}[t]
    \centering
    \begin{minipage}{0.75\textwidth}
    \begin{minipage}[t]{0.48\linewidth}
        \centering
        \includegraphics[width=\linewidth]{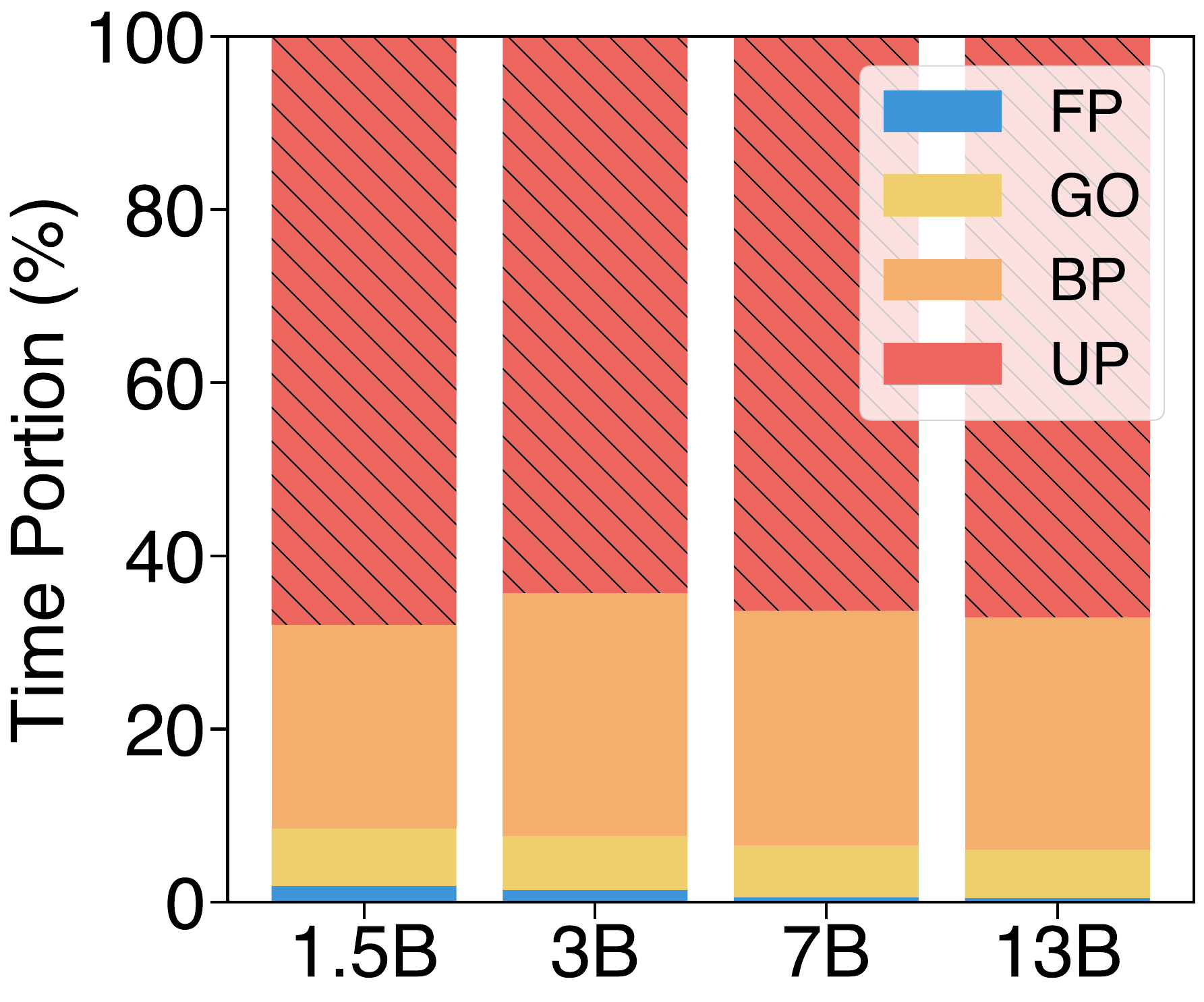} 
        \caption{Per-iteration time breakdown under ZeRO-Offload when training Qwen2.5-\{1.5B,3B\} and Llama2-\{7B, 13B\} models.
        }   
        \label{fig:baseline_iteration_breakdown} 
        \Description{Breakdown of iteration time. }
    \end{minipage}\hfill
    \begin{minipage}[t]{0.48\linewidth}
        \centering
        \includegraphics[width=\linewidth]{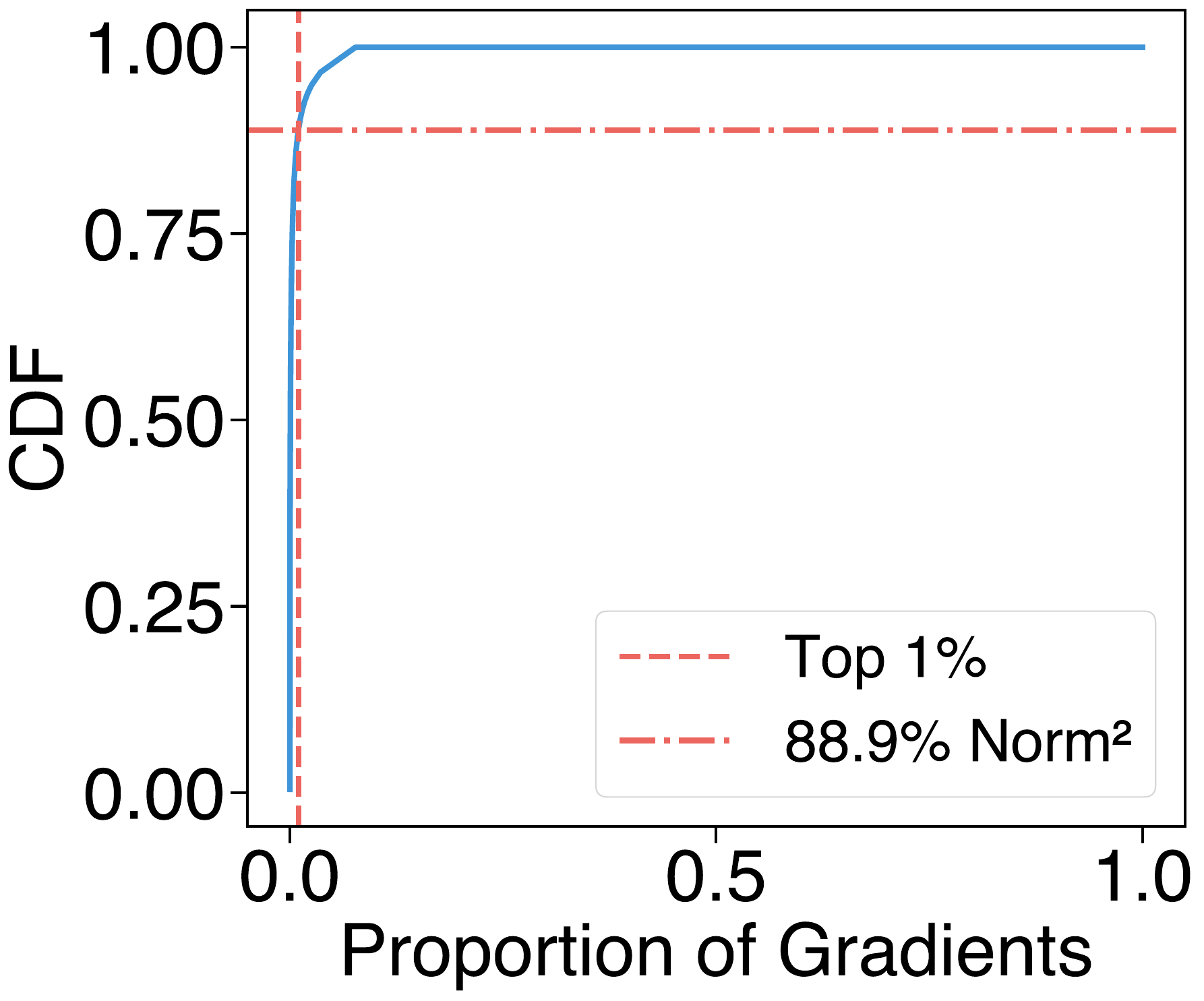}
        \caption{CDF of gradient norm squared across all gradients on fine-tuning Qwen2.5-0.5B on the \texttt{Alpaca52K} dataset.
        }
        \label{fig:grad_norm_cdf}
        \Description{CDF of gradient norm. }
    \end{minipage}
    \end{minipage}
\end{figure}

\subsection{Memory-Efficient Training}
\label{subsec:memory_efficient_training}

Training memory is dominated by persistent states (i.e., parameters, gradients, and optimizer states), while activations are short-lived. In \texttt{BF16}/\texttt{FP16}, if parameters take $M$, gradients add another $M$, and optimizer states add $2M$, approximately $4M$ in total (see \tref{tab:offload_example}). For Llama2-7B, that is 14GB each for parameters and gradients, and 28GB for optimizer states—totaling 56GB, which easily exceeds the memory limit of an A100 40GB GPU.

\phead{Offloaded Training.} To mitigate the GPU memory bottleneck, some offloading techniques~\cite{ren2021zero_offload, zeroinfinity_sc21, swapadvisor_asplos2020, sentinel_hpca2021} have been proposed to transfer model parameters and optimizer states from the GPU memory to other storage, such as CPU DRAM or NVMe storage. Those methods reduce the GPU memory consumption and enable ``out-of-core'' training of larger models. ZeRO-Offload~\cite{ren2021zero_offload} in the DeepSpeed ZeRO series~\cite{ren2021zero_offload,rajbhandari2020zero,zeroinfinity_sc21}, is one of those state-of-the-art systems.

\fref{fig:offload_comparision}(a) illustrates the workflow of ZeRO-Offload. In each training iteration (\text{${Step}_I$}), the forward pass (\texttt{${FP}_I$}) and backward pass (\texttt{${BP}_I$}) are executed on the GPU. During the backward pass, the gradients are computed and then transferred from GPU to CPU memory, where the CPU performs the parameter update (${UP}_I$). Once the update is complete, the updated parameters $\Delta W_I$ are fetched back to the GPU for the forward pass in the next training iteration. During the \texttt{UP} stage (the parameter updating), the GPU is idle, waiting for the updated parameters to be fetched from the CPU.

\phead{Parameter-Efficient Fine-Tuning (PEFT).} As summarized in \tref{tab:method_comparison}, PEFT methods reduce training cost by updating only a small subset of parameters instead of all parameters. Typical PEFT approaches include adapters~\cite{sparseadapter_arxiv2022}, prefix-tuning~\cite{prefix_tuning_arxiv2021}, LoRA~\cite{lora_iclr2022, qlora_nips2023}, and GaLore~\cite{galore_2024}. Among them, LoRA is the most widely used: it trains low-rank adapters while freezing the original weights, updating less than 1\% of parameters. This design enables low memory usage and fast training. However, this efficiency comes at the cost of representational fidelity. PEFT methods are structurally incapable of full-parameter adaptation, which can lead to suboptimal convergence and fail to capture the deep weight shifts required for domain specialization~\cite{revisiting_acl2022, parameter_acl2023, parameter_2025arxiv, dai2024discovering, patel2025guide}. In this paper, we target full fine-tuning to maximize generality. Nevertheless, ZenFlow remains compatible with PEFT methods and can be combined when parameter efficiency is required.


\subsection{Problems and Insights} 
\label{subsec:problem_insight}

\phead{Problem: Fast GPU execution is frequently stalled by slow CPU-side update.} To understand the performance bottleneck of offloading-based fine-tuning, we break down per-iteration time when fine-tuning Qwen2.5 and Llama2 model series with various sizes (see \fref{fig:baseline_iteration_breakdown}). We use ZeRO-Offload with fully parallelized \texttt{CPUAdam} optimizer.

Each iteration is broken down into four stages: forward pass (\texttt{FP}), backward pass (\texttt{BP}), gradient offloading (\texttt{GO}), and parameter update (\texttt{UP}). 
With ZeRO-Offload, the forward and backward passes run on the GPU, the gradients are transferred to the CPU, and the update phase is computed entirely on the CPU with the \texttt{CPUAdam} optimizer.  
Despite parallelizing the \texttt{CPUAdam} optimizer across 128 CPU threads, the CPU-side update is a significant bottleneck (see \fref{fig:baseline_iteration_breakdown}). For example, when fine-tuning a Llama2-7B model, the update stage on the CPU takes $\sim$4,600~$ms$---over twice as long as the backward pass time of 2,000~$ms$, which leads to GPU idling, as the GPU must wait for the CPU to complete its update computation before proceeding to the next iteration.  

To improve GPU utilization, one natural solution from StrongHold~\cite{stronghold_sc22} is to overlap CPU updates with GPU computation using layer-wise scheduling 
(\fref{fig:offload_comparision}~(b)). In this strategy, each layer's gradients are offloaded and updated sequentially during the backward pass, allowing the CPU to begin updating earlier layers while the GPU continues computing later ones. This pipelined execution enables partial overlap between CPU-side 
updates and the GPU-side backward computation. For example, as soon as the gradient for layer $l_n$ is computed, it can be offloaded and the corresponding update initiated on the CPU, rather than waiting for the full backward pass (for all remaining layers from $l_{n-1}$ to $l_0$) to complete.
This allows updated parameters to be uploaded back to the GPU earlier, reducing GPU idle time. 

Although layer-wise scheduling can partially overlap CPU and GPU tasks, the CPU update phase is too long to be fully hidden (4,600~$ms$ on the CPU vs. 2,000~$ms$ on the GPU).  

As a result, \emph{GPU execution is frequently stalled, waiting for the CPU update even with aggressive multi-threaded optimization and optimal overlap.}  

Furthermore, parameter upload introduces additional stalls. While gradient offloading overlaps naturally with the backward pass (since gradients can be streamed asynchronously to CPU), the next iteration must wait for updated parameters to return from CPU via PCIe. This adds additional 500~$ms$ stall per iteration as shown in \fref{fig:offload_comparision}(c).

\begin{table}[t]
    \centering
    \caption{Resource requirement for fine-tuning Llama2-7B (\texttt{BF16}) on $4\times$ A100 80GB GPUs with ZeRO-Offload.}
    \label{tab:offload_example}
    \begin{adjustbox}{width=0.6\textwidth}
    \begin{tabular}{lccc}
    \toprule
    \multirow{2}{*}{\textbf{Memory}} 
    & \textbf{Param.} & \textbf{Optim. States} & \textbf{Gradient} \\
    & 14GB ($M$) & 28GB ($2M$) & 14GB ($M$) \\
    \midrule
    \multirow{2}{*}{\textbf{Computation}} 
    & \textbf{FP on GPU} & \textbf{BP on GPU} & \textbf{UP on CPU} \\
    & 45$ms$/step & 2,000$ms$/step & 4,600$ms$/step \\
    \midrule
    \multirow{3}{*}{\textbf{Communication}} 
    & \makecell{\textbf{CPU-GPU} \\ \textbf{Bandwidth}} 
    & \makecell{\textbf{Gradient} \\ \textbf{Accumulation}} 
    & \makecell{\textbf{Param.} \\ \textbf{Update}} \\
    & \makecell{$\sim$28GB/s}  
    & \makecell{GPU$\rightarrow$CPU \\ 14GB ($M$)} 
    & \makecell{CPU$\rightarrow$GPU \\ 14GB ($M$)} \\
    \bottomrule
    \end{tabular}
    \end{adjustbox}
\end{table}

Next, we present insights to address the core limitation of existing offloaded training systems.

\phead{Insight \#1: Gradients/parameters with different importance should be treated differently to decouple GPU/ CPU execution and reduce communication.} 
It has been 
observed that the gradients of different parameters have different importance in deep neural network  training~\cite{aji2017sparse,strom2015scalable,lin2017dgc,dryden2016communication}. For example, \fref{fig:grad_norm_cdf} shows that top 1\% of gradients account for $\sim$90\% of the gradient norm. However, the current offloading techniques \emph{overlook} this difference in gradients, treating them equally and offloading all of them to the CPU regardless of their impacts. This uniform treatment introduces unnecessary inefficiencies. 
Take Llama2-7B as an example. An important subset of its parameters could be updated and transferred in $\sim$50~$ms$, but under the uniform scheme they stall until the full CPU update (4,600~$ms$) and parameter transfer (500~$ms$) complete. In other words, the few important parameters that drive learning are forced to wait through $\sim$5~$s$ of unnecessary work dominated by unimportant ones. 

This observation leads to a key research question: \emph{Can we decouple important and less important gradients, and assign them to different hardware resources accordingly?} To mitigate unnecessary stalls, we propose updating important gradients directly on the GPU---leveraging its high computing bandwidth---while offloading and accumulating the remaining less important gradients on the CPU. This design relaxes the tight coupling between GPU execution and the full CPU update cycle, allowing the GPU to proceed without waiting on low-priority updates. The CPU-side accumulation proceeds \emph{concurrently} and is typically fast enough (e.g., $\sim$500~$ms$) to be hidden within the backward pass, thereby avoiding additional delay and improving GPU utilization.

\phead{Challenge: Full gradient view is expensive in fully sharded distributed training.}
To leverage gradient importance during training, selecting important gradients correctly and efficiently is a critical step. 

A simple yet effective approach is top-$k$ selection, which retains the gradients with the highest magnitudes. This strategy has been widely studied and applied in prior work \cite{aji2017sparse,maskllm_nips2024,huang2025castcontinuousdifferentiablesemistructured,jang2024smartinfinityfastlargelanguage}. However, the top-$k$ selection assumes access to the full set of gradients---what we refer to as a \emph{global gradient view}.

In fully sharded distributed training, this assumption breaks. Each GPU holds only a fraction of the model parameters and computes gradients for its local shard, making global top-$k$ selection \emph{expensive}, because constructing a fully global view would require gathering and synchronizing gradients across all devices, incurring significant communication overhead and causing peak memory usage spikes due to gathered full gradient matrix. 
Therefore, directly applying existing global top-$k$ selection is not practical.   


\begin{figure}[!t]
    \centering
    \includegraphics[width=0.75\textwidth]{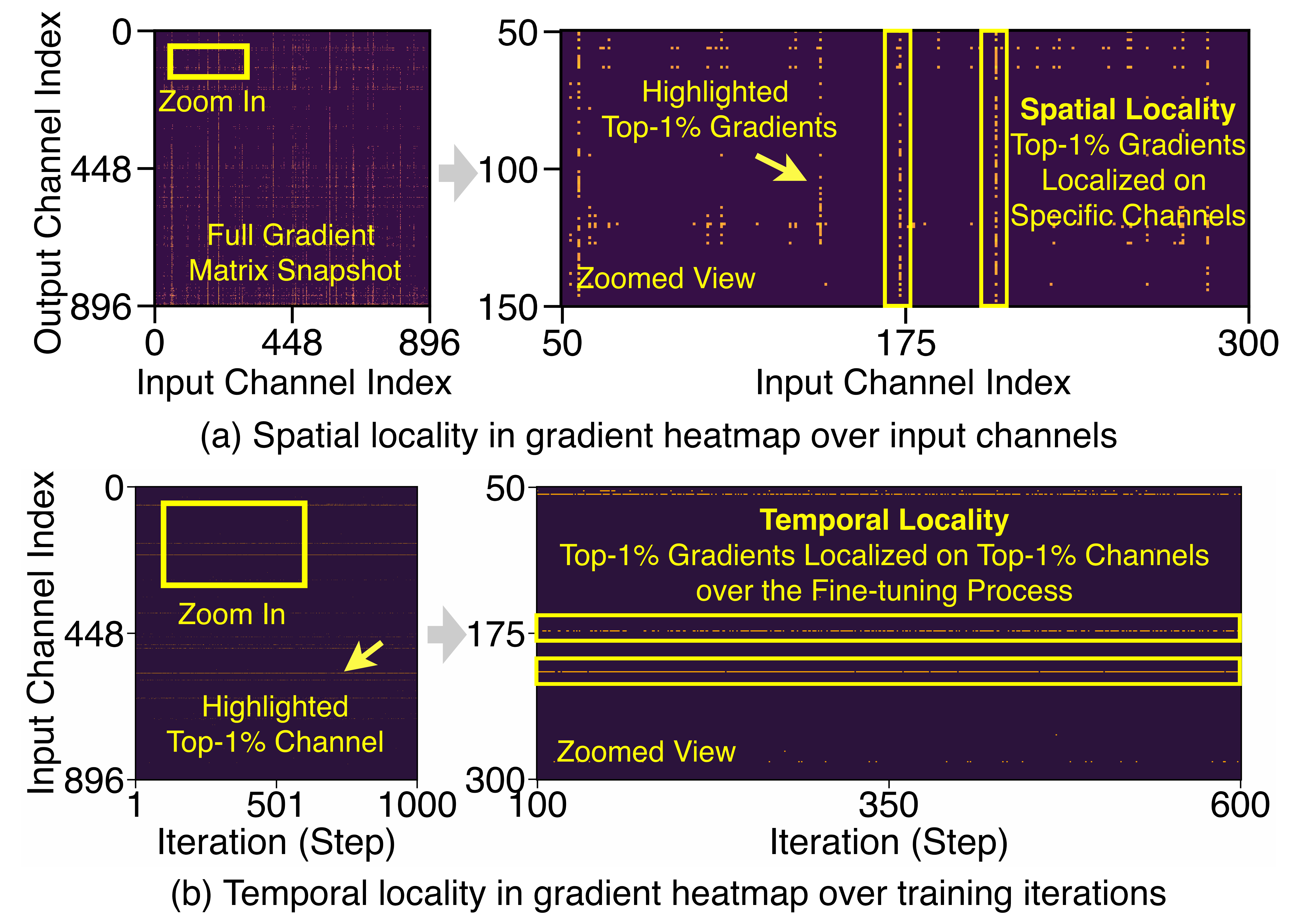}
    \caption{Localities of gradients during fine-tuning. 
    } 
    \Description{Spatial-temporal locality heatmap showing the top-1\% gradients.}
    \label{fig:gradient_heatmap}
\end{figure}

\phead{Insight \#2: Important LLM gradients show both spatial and temporal locality.} To address the challenge posed by the lack of a global gradient view in fully sharded distributed training, we next investigate an important research question: \emph{Do important gradients in LLM fine-tuning exhibit spatial or temporal locality that we can exploit to approximate global top-$k$ selection more efficiently?}

To answer this, we analyze the gradient importance distribution during fine-tuning. As a representative case,  we use Qwen2.5-0.5B on the \texttt{Alpaca52K} dataset to illustrate. We also evaluate whether this locality generalizes across model architectures and optimizers in \sref{subsec:eval_gradient_locality}.

First, we observe clear \textbf{spatial locality} in the gradient distribution. As shown in \fref{fig:gradient_heatmap}(a), the top 1\% of gradients (highlighted in orange) are not uniformly distributed but instead concentrated in a narrow subset of columns, each corresponding to a specific input channel. This pattern persists across iterations 
(see \fref{fig:gradient_heatmap}(b) and discussed next).

This result suggests that a small set of input features consistently receive large updates during fine-tuning. 
This is because transformer activations are often localized across channels during inference~\cite{jaszczur2021sparse,shen2020powernorm}, and such patterns propagate backward to the gradients.

Based on this, instead of performing the expensive global top-$k$ selection over the entire parameter matrix, 
we can approximate gradient importance by identifying and tracking a small set of important input channels---an approach significantly cheaper in both computation and communication.

Second, we find that these important input channels also exhibit \textbf{temporal locality}. As shown in \fref{fig:gradient_heatmap}(b), the top 1\% of important input-channel indices remain stable across iterations, forming persistent horizontal bands over time. This indicates that the same small subset of input channels consistently receives large updates during fine-tuning: these channels contribute more to the learning process than others, and their importance remains stable over time.
While occasional deviations occur---likely due to exploration of alternative subspaces---the overall importance remains stable. 
Empirically, the top-1\% channels exhibit a retention rate above 90\%, while the retention rate approaches 100\% under a looser threshold (e.g., top-5\%). Therefore, we can efficiently track important gradients by maintaining a stable set of channels.

This spatio-temporal locality enables a lightweight approximation to global top-$k$ selection: rather than recomputing gradient importance every iteration, we can approximate important gradients efficiently using 
a slowly-updated set of important channels.  
This drastically reduces communication and synchronization overhead, while preserving high fidelity in identifying important updates. 

%% file: contents/overview.tex
\section{\system Design}
\label{sec:overview}

The challenges and insights discussed in \cref{sec:background} motivate \system, a system that leverages gradient importance and gradients' spatio-temporal locality to mitigate the I/O bottleneck and GPU stalls in offloading training. \system decouples gradient updates across heterogeneous hardware and minimizes communication overhead without sacrificing accuracy. 
This section presents the design of \system with the following goals: 

\noindent$\bullet$~\phead{Goal \#1.} Minimize GPU stalls caused by slow CPU (i.e., I/O and computation overhead associated with unimportant gradients and parameters).

\noindent$\bullet$~\phead{Goal \#2.} Ensure training reproducibility with deterministically scheduled update process. 

\noindent$\bullet$~\phead{Goal \#3.} Guarantee training accuracy by preserving all (important and less important) gradient information.

\subsection{Importance-aware Offloading Workflow}
\label{subsec:async_offloading} 

We begin by presenting the overall workflow of \system to provide a high-level view of how it decouples gradient updates across GPU and CPU. In this subsection, we assume that an important subset of gradients and their corresponding parameters have already been identified. The mechanism for efficiently selecting this subset is described later in \sref{subsec:gradient_selection}. 
Here, we focus on how \system manages the importance-aware offloading and update process once the selection is in place. 

\phead{Important updates: fast GPU path.} As shown in \fref{fig:offload_comparision}(d), \system assigns the important gradient and parameters (highlighted in purple) to the GPU, while offloading the unimportant ones (shown in gray) to the CPU. 
At each iteration, \system performs a standard forward and backward pass with the full set of parameters. 
Once gradients are computed, the pre-identified important gradients remain on the GPU, where a \textit{selective-optimizer}, initialized only with corresponding parameter subsets, performs an \emph{in-place} update.  
This GPU-side update is lightweight, as it operates on a small subset of parameters, and it completes on the GPU without introducing stalls between iterations. 

\phead{Unimportant updates: slow CPU path.} Less important gradients are offloaded to the CPU and gradually accumulated over several iterations. 
These gradients are \emph{not} discarded. They are simply delayed until they become important enough---a process of what we call \emph{gradient accumulation}.  
Once the accumulated gradients become large enough to matter, {\system} performs a full parameter update on the CPU. In practice, this trigger is implemented via a fixed accumulation interval or an adaptive threshold (\sref{subsec:zero_bubble_pipeline}, \sref{subsec:sensitivity}), ensuring bounded staleness.
This CPU-side update, though being slow, is carefully scheduled, so that it 
(1)~perfectly overlaps with GPU computation to avoid stalls 
(see \sref{subsec:zero_bubble_pipeline}),
and (2)~ensures training reproducibility and accuracy (see \sref{sec:accuracy_guarantee}).  

In effect, \system updates important parameters more frequently using \emph{fast} GPU, while updating less important ones less often on \emph{slow} CPU, but with full gradient information.

This design achieves performance gains from three parts: 
(1)~In each iteration, only a small subset of parameters are updated efficiently in place on GPU, and these updates are independent from  the slow CPU-side.
(2)~The slow and compute-intensive CPU updates for large unimportant gradients are fully overlapped with multiple iterations of GPU computation, effectively hiding their latency. 
(3)~Update I/O overhead for unimportant parameters is amortized as {\system} only updates and transfers them to GPU when they become large enough. 
Together, these optimizations minimize stalls. 

\subsection{Zero-stall Pipeline}
\label{subsec:zero_bubble_pipeline}

While the importance-aware offloading design has the potential to minimize GPU stalls by selectively updating a small set of parameters, the CPU-side update can still become a bottleneck if not carefully overlapped with GPU computation. 
Moreover, naïvely creating an extra \textit{selective-optimizer} on the GPU may incur unnecessary memory overhead. 
We next describe how \system addresses both challenges, by hiding CPU update latency and managing memory efficiently, to realize a truly zero-stall pipeline.

\phead{Modeling I/O Efficiency.} We now provide an analytical comparison of the offloading workflow I/O traffic between \system and DeepSpeed ZeRO-Offload, demonstrating that \system{}'s pipeline significantly reduces communication overhead. For this analysis, we consider the case of fine-tuning with half-precision (\texttt{BF16}/\texttt{FP16}).
In each iteration, ZeRO-Offload transfers the gradient generated from the GPU to the CPU, equivalent to one model copy ($M$). 
After the CPU optimizer completes the update, the updated parameters are transferred back to the GPU---another model copy ($M$). Therefore, the total I/O traffic per iteration is $2M$.

In contrast, \system offloads only the gradients for less important parameters. Let $k$ denote the top-$k$ ratio (i.e., the fraction of gradients considered important) and $S$ the number of accumulation rounds for the unimportant gradients. In each iteration, \system transfers only the $(1-k)\cdot M$ unimportant gradients to the CPU. After $S$ iterations, the CPU performs a parameter update and sends back the corresponding $(1-k)\cdot M$ updated parameters. Therefore, the average I/O traffic per iteration in \system is: $\frac{(S+1)\cdot(1-k)\cdot M}{S}$. 

Take $S=4$ (one representative configuration of the accumulation rounds from our analysis for hiding CPU-side updates we will discuss shortly), and $k=0.1$ as an example, the average I/O traffic per iteration becomes $1.125M$, which is nearly a $2\times$ reduction compared to 
ZeRO-Offload.

\begin{figure}[t]
    \centering
    \includegraphics[width=0.75\textwidth]{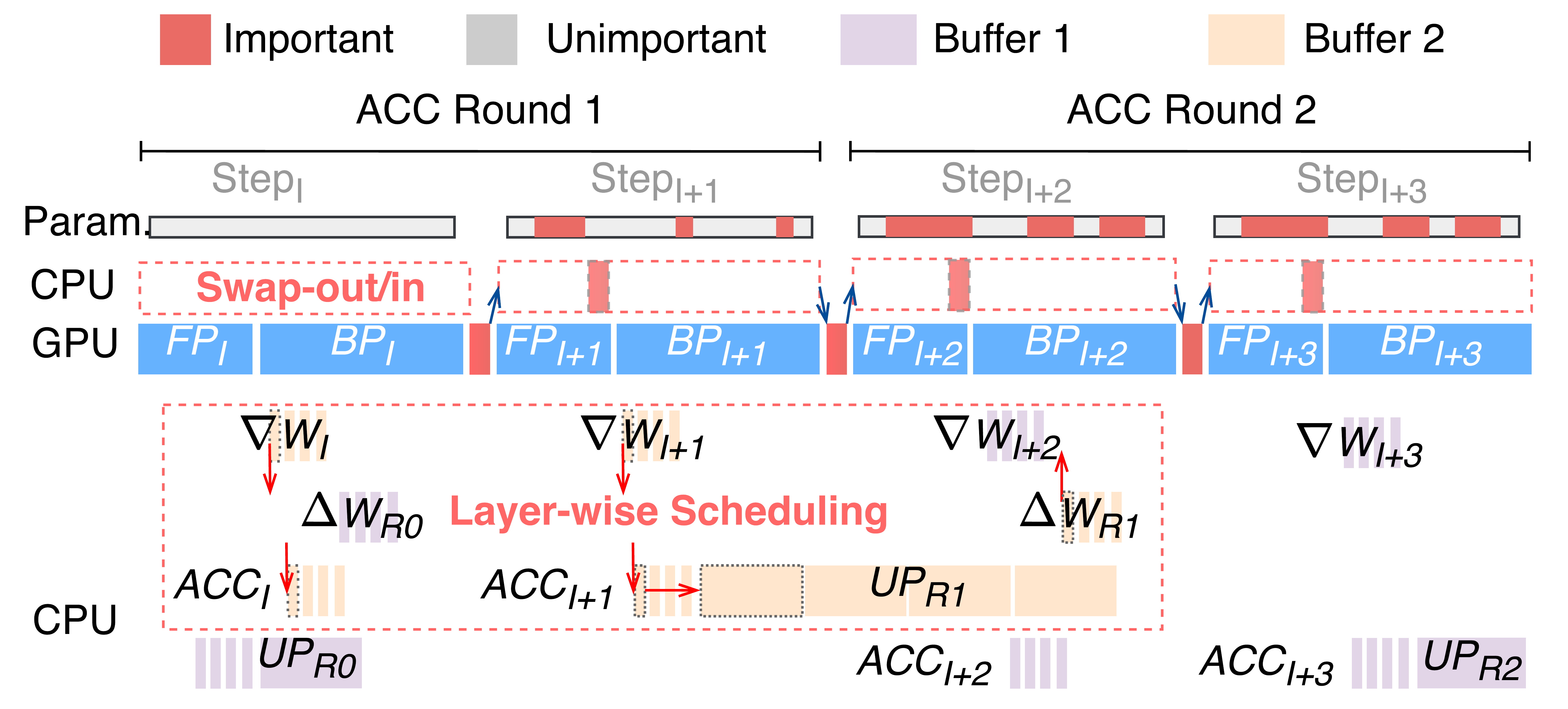}
    \caption{Zero-stall pipelining with double buffering. \system overlaps GPU computation with CPU-side gradient accumulation and updates.
Across training steps, important gradients are updated immediately on the GPU, while unimportant gradients are offloaded and accumulated on the CPU.
The CPU maintains two buffers and alternates between accumulation (\texttt{ACC}) and update (\texttt{UP}) phases across rounds.
While one buffer accumulates gradients from the current iterations, the other buffer applies parameter updates in the background.
Layer-wise scheduling and swap-out/in ensure that CPU updates and parameter transfers are fully overlapped with GPU \texttt{FP}/\texttt{BP}, preventing GPU stalls.}
    \label{fig:zero_bubble_pipeline}
    \Description{Zero-bubble pipeline.}
\end{figure}

\phead{Hiding CPU-side Updates.}
In addition to reducing I/O traffic, \system also hides CPU-side update latency by overlapping it with the GPU computation. Our profiling shows that, for a Llama2-7B model, CPU update latency is approximately 4,600~$ms$ with 128 CPU threads (fully parallelized on our GPU node) and $\sim$6,200~$ms$ with limited resources (e.g., with only 32 CPU threads). 
Uploading updated parameters to the GPU typically takes $\sim$500~$ms$. 
In contrast, the GPU forward pass ($\sim$45~$ms$) and backward pass ($\sim$2,000~$ms$) take around $\sim$2,045~$ms$. This means the CPU update can be overlapped with $2.3\times$ to $3.1\times$ the GPU compute time.

Critically, we observe that unimportant gradients usually require 4-6 accumulation steps before crossing the importance threshold for update (see \sref{subsec:sensitivity}). This enables their CPU-side updates to be effectively masked by 4-6 iterations of GPU forward/backward computation, achieving near-complete overlap without additional delay. Empirically, we set the accumulation interval to 4 steps ($S=4$), which is sufficient to hide CPU update latency in most cases while bounding staleness (see \sref{sec:accuracy_guarantee} for theoretical analysis).

To achieve this, \system employs a double-buffering strategy on the CPU. It maintains two gradient buffers: one for concurrently accumulating gradients from the current iteration, and one for holding previously accumulated gradients used for parameter updates. While the GPU executes the forward and backward passes and sends new gradients, the CPU concurrently applies updates accumulated in one buffer. See the step of $\textit{UP}_{R1}$ in \fref{fig:zero_bubble_pipeline}, where parameters are updated using gradients accumulated from both $\textit{Step}_I$ and $\textit{Step}_{I+1}$. 

Once the update is complete, the buffers are swapped, and the buffer used for parameter updates is cleared in order to store gradients in the next accumulation cycle. 
This design allows CPU updates to be completely hidden and run transparently in the background, fully overlapping with GPU training without introducing stalls.

\phead{Swapping out/in and Layer-wise Scheduling.} 
To avoid excessive GPU memory consumption from the GPU-side \textit{selective-optimizer}, \system swaps out its optimizer states to CPU and swaps back in before next update on GPU. These states are relatively small and can be transferred efficiently. To further reduce memory cost, swapping is performed in a layer-wise manner, ensuring that only one layer's optimizer state resides on GPU at any time. This design maintains high memory efficiency by preventing additional overhead beyond what is required for a single layer. The same layer-wise scheduling is applied to gradient offloading and CPU-side updates to fully overlap communication and computation.

\phead{Hyperparameter Auto-tuning.} To fully leverage the locality of important gradients while maintaining accuracy, we design \systemauto, which adaptively tunes the update interval based on observed learning dynamics. Specifically, \systemauto monitors gradient changes across GPUs using a lightweight coordination proxy (detailed in \sref{subsec:gradient_selection} and \sref{subsec:sensitivity}). 
For unimportant gradient part, \systemauto tracks the average accumulated channel gradient norm and compares it to the average one of the important part. Once the unimportant gradient part becomes comparable to important ones, \systemauto immediately triggers its CPU-side update, ensuring timely parameter refresh and stable convergence.

\subsection{Gradient Selection}
\label{subsec:gradient_selection}

We now turn to the key question: \emph{how to select important gradients for GPU-side updates in a scalable, low-overhead manner.}   
A natural approach is to prioritize gradients with large magnitudes, as they typically contribute more to learning. However, in fully sharded training, computing and comparing all gradient norms globally is 
expensive. To make this feasible in distributed training, \system introduces a lightweight yet effective approximation that leverages the spatio-temporal 
locality of important gradients discussed in \sref{subsec:problem_insight}.

\begin{figure}[t]
    \centering
    \includegraphics[width=0.75\textwidth]{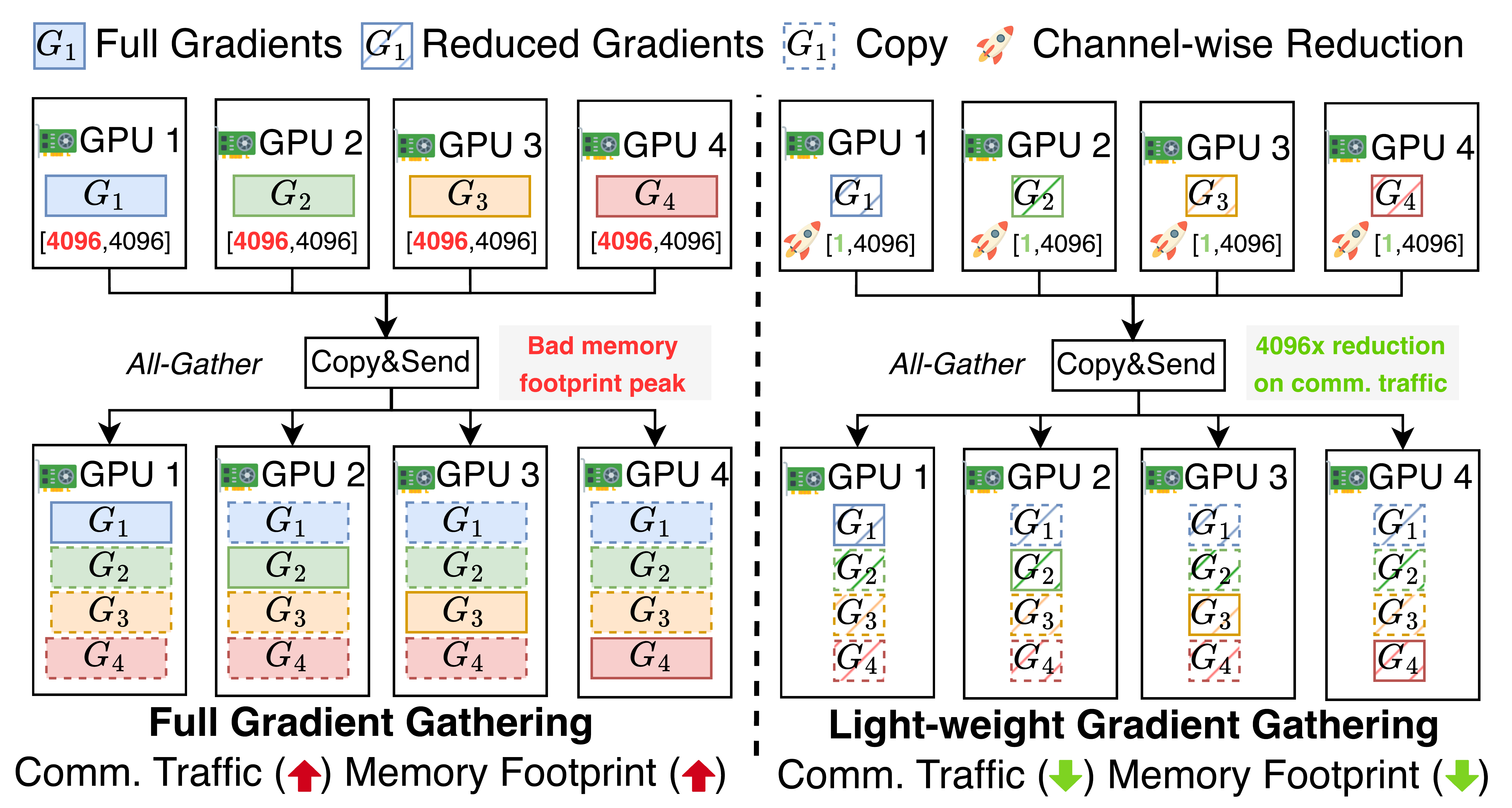}
    \caption{Gradient gathering strategies. \textbf{Left:} Full gradient gathering collects and all-gathers full gradient matrices across GPUs, incurring high communication volume and large peak memory footprint.
\textbf{Right:} \system performs light-weight gradient gathering by first reducing gradients along the channel dimension and exchanging only compact per-channel summaries.} 
    \label{fig:gradient_gathering}
    \Description{Gradient gathering strategies.}
\end{figure}

\phead{Lightweight Proxy for Gradient Ranking.} 
In fully sharded training, selecting important gradients across GPUs poses a major communication challenge. A straightforward solution is to rank all gradients globally by magnitude and prioritize the top-$k$ for updates. However, this requires collecting all gradients across devices via expensive \texttt{AllGather} .

For example, consider fine-tuning Llama2-7B across 4 GPUs shown in \fref{fig:gradient_gathering}(left). A naïve global ranking would require exchanging tens of gigabytes of gradients per iteration, incurring significant communication overhead. 
Even a single weight matrix imposes non-trivial communication overhead. Consider \texttt{q\_proj.weight} in a transformer layer, shaped $[4096, 4096]$, containing 16.8M parameters. In a 4-way sharded setup, each device holds a $[1024, 4096]$ partition with 4.2M parameters. Using \texttt{BF16/FP16}, this amounts to 8MB per GPU. Aggregating gradients for this matrix across all devices requires 96MB of data transfer per iteration. Extrapolating to a 7B-parameter model, this results in a total gradient communication volume of 40GB per iteration.

To address this bottleneck, \system employs a communication efficient proxy: instead of gathering full $(n \times m)$ gradients, where $n$ is the output dimension and $m$ is the input dimension of a weight matrix, each GPU computes and shares per-column gradient norms squared (i.e., the sum of squared gradient values within each column). This reduces both communication and top-$k$ selection complexity from $O(nm)$ to $O(m)$ while preserving gradient magnitude information and omitting only directional components. For example, consider a \texttt{q\_proj.weight} from Llama2-7B with shape $4096 \times 4096$. Rather than transferring the full 32MB gradient matrix in BF16, each GPU shares a 4096-dimensional vector (8KB), reducing communication volume by over 4,000$\times$ with negligible impact on importance estimation (\fref{fig:gradient_gathering}(right)).

\begin{figure}[t]
    \centering
    
    \includegraphics[width=0.75\textwidth]{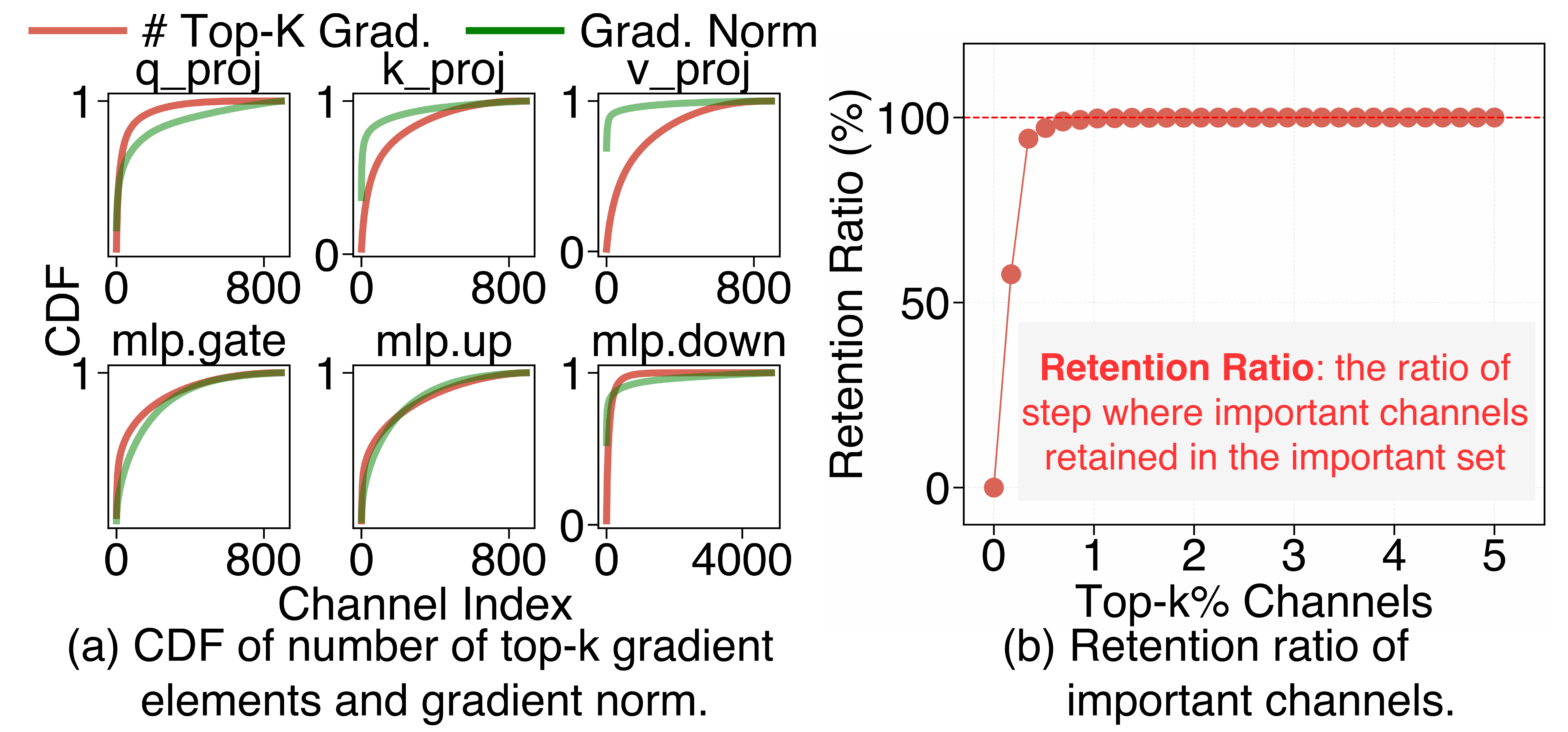}
    \caption{CDF of top-$k$ elements and gradient norm. The channel index is sorted by the number of top-$k$ elements.}
    \Description{CDF of top-k elements and gradient norm.}
    \label{fig:gradient_selection} 
\end{figure}

\phead{Spatial Locality.}  
We empirically validate the effectiveness of the approximation. \fref{fig:gradient_selection}(a) shows the cumulative distribution of top-$k$ gradients and per-channel gradient norms during Qwen2.5-0.5B fine-tuning on \texttt{Alpaca52K}. Channels are sorted by how frequently they contain top-$k$ gradients. The results show that 60-90\% of top-$k$ gradients are concentrated in just the top 10\% of channels, and that per-channel gradient norms are strongly correlated with this top-$k$ density. This confirms that channel-level summaries are a reliable indicator of gradient importance.

\phead{Temporal Locality.} 
To further reduce the selection overhead, we examine whether important channels persist across iterations. \fref{fig:gradient_selection}(b) reports the retention ratio—the fraction of steps where previously selected channels remain within the important set. We find that once the top-1\% of channels are tracked, the retention ratio already approaches 100\%, and stays nearly perfect as the threshold widens to 3-5\%. This strong temporal locality shows that caching and reusing channel indices is sufficient to cover important updates, avoiding repeated computation. In multi-GPU settings, this further reduces cross-device coordination and amortizes selection cost, enabling scalable importance-aware training.

%% file: contents/accuracy.tex
\section{Accuracy and Reproducibility Guarantee}
\label{sec:accuracy_guarantee}

To ensure that {\system} training delivers robust quality, we explicitly address the requirements of convergence and reproducibility across runs. 

\subsection{Convergence.} We now show that \system{}'s asynchronous  offloading design does not affect the convergence property of existing optimizers. 

Under standard assumptions commonly used in asynchronous optimization analysis~\cite{staleness_arxiv2018,asynchronous_sgd_nips2015,zhang2015staleness,convergence_PMLR2016}, we show that \system preserves the convergence rate of \(\mathcal{O}(1/\sqrt{T})\) with a bounded staleness factor where $T$ is the total number of iterations. Such a convergence rate is the same as the ideal rate of synchronous SGD~\cite{staleness_arxiv2018,zhang2015staleness}. 

\phead{Why prioritizing high-norm channels.} Prioritizing high-norm gradient channels is consistent with standard descent analysis. For an $L$-smooth objective, one SGD step satisfies
\[
f(w_{t+1}) \leq f(w_t) - \eta \langle \nabla f(w_t), g_t\rangle + \frac{L\eta^2}{2}\|g_t\|_2^2.
\]
Channels with larger gradient norm therefore have disproportionate influence on the descent direction. Existing top-$K$ sparsification results further show that retaining large-norm channels with error compensation can preserve convergence behavior~\cite{alistarh2018convergence,stich2018sparsified}.

\phead{Partial staleness in asynchronous training.}  We consider a mixed asynchronous training setup where the parameters are partitioned into two disjoint sets: \( \theta = [\theta^{(g)}, \theta^{(c)}] \). The gradients w.r.t.\ \( \theta^{(g)} \) are computed and applied \emph{immediately} on the GPU every iteration, while the gradients w.r.t.\ \( \theta^{(c)} \) are \emph{accumulated on the CPU} over \( S \) iterations (typically \( S=4 \)) and then applied synchronously.  
Formally, at each training step \( t \), we have the following. 
\[
\theta_{t+1}^{(g)} = \theta_t^{(g)} - \alpha_t \nabla_{\theta^{(g)}} L(\theta_t)
\]
\[
\theta_{t+1}^{(c)} =
\begin{cases}
\theta_t^{(c)} - \alpha_t \cdot \frac{1}{S} \sum_{i=t-S+1}^{t} \nabla_{\theta^{(c)}} L(\theta_i), & \text{if } t \bmod S = 0 \\
\theta_t^{(c)}, & \text{otherwise}
\end{cases}
\]
where $ \alpha_t $ is the learning rate and $ L(\theta) $ is the objective loss function. The GPU handles the forward and backward passes for both $ \theta^{(g)} $ and $ \theta^{(c)} $, but only updates $ \theta^{(g)} $ immediately. The CPU accumulates the gradients for $ \theta^{(c)} $ over $ S $ iterations before applying the update.

Next, we model a partially stale update system with a bounded delay. Let
$ \rho = \frac{\sup_t\mathbb{E}[\|\nabla_{\theta^{(c)}}L(\theta_t)\|_2^2]}{\sup_t\mathbb{E}[\|\nabla L(\theta_t)\|_2^2]} $,
representing the fraction of total gradient-norm energy that resides in the delayed coordinates (on the CPU). Empirically, we observe $ \rho \approx 0.10 $, as the GPU handles 90\% of the gradient energy.

\phead{Bounded-staleness result.} With common assumptions of unbiased stochastic gradients, bounded variance, and Lipschitz smoothness~\cite{koloskova2022sharper,asynchronous_sgd_nips2015,convergence_PMLR2016,staleness_arxiv2018}, and letting \(\rho\) denote the fraction of gradient energy at the
CPU side, we have the following: 
\[
\frac{1}{T}\sum_{t=1}^{T}\mathbb{E}\!\left[\|\nabla L(\theta_t)\|_2^{2}\right]
\;\le\;
\mathcal{O}\!\Bigl(\sqrt{\tfrac{1+\rho S}{T}}\Bigr).
\]
The term \(\mathcal{O}\Bigl(\sqrt{\tfrac{1}{T}}\Bigr)\) is the standard SGD rate; the factor \(\sqrt{1+\rho S}\)  quantifies the extra cost of staleness. With \(S=4\) and \(\rho\approx0.10\), this factor is
\(\sqrt{1.4}\approx1.18\), i.e., an 18 \% slowdown relative to ideal
synchronous SGD.

\phead{Warmup mitigates early-stage staleness.} The convergence bound under the partial staleness includes a penalty term with the form $\sqrt{1 + \rho S}$, where \(\rho\) is the fraction of gradient-norm energy in the delayed coordinates and \(S\) is the accumulation interval. This bound assumes uniform gradient energy across training steps.
However, in practice, the gradients are \emph{not} uniformly distributed. Early training steps contribute disproportionately to optimization progress, as the gradient norms are significantly larger during this phase. Empirical and theoretical studies suggest that the gradient energy often decays as \(\mathbb{E}[\|\nabla L(\theta_t)\|^2] \sim t^{-\beta}\), with \((0 < \beta < 1)\)~\cite{luo2025multi,hoffmann2022training,kaplan2020scaling}.

\system exploits this observation by applying synchronous updates (i.e., no staleness) during the initial \(\tau\) warmup steps, and then switches to asynchronous offloading for the remaining \(T - \tau\) steps. 
This strategy eliminates staleness where it causes the most harm, while preserving efficiency later when gradients are smaller and more stable.

To quantify the effect, we compute the \emph{gradient-weighted} penalty using a continuous approximation of the \(p\)-series.
\[
\text{Penalty}(\beta)
\approx
\sqrt{1 + \rho S \cdot \left(1 - \left(\frac{\tau}{T}\right)^{1 - \beta}\right)}.
\]
This closed form captures the diminishing impact of the delayed gradients as training progresses.

For example, with a 5\% warmup (\(\tau/T = 0.05\)), \(S = 4\), \(\rho = 0.1\), and \(\beta = 0.6\) (typically ranging from 0.4 to 0.6~\cite{kaplan2020scaling,luo2025multi}), the multiplicative penalty factor decreases from approximately 1.18 without warmup to 1.13 with warmup---equivalently, the excess penalty decreases from 18.3\% to 13.1\%.

Thus, the bound for \system incurs only about 13\% overhead relative to the ideal SGD rate. Modern optimizers such as Adam and AdamW may further mitigate gradient staleness through momentum and adaptive learning rates~\cite{lin2017dgc,adam_2017,shazeer2018adafactor,loshchilov2017decoupled}. With this modest theoretical overhead, \system achieves up to 5$\times$ end-to-end speedup, a 2$\times$ reduction in I/O traffic, and near-zero stalls. We present the end-to-end speedup and a detailed breakdown of the performance gains in \sref{sec:evaluation}.

We use SGD in the derivation because it provides a general and transparent explanation of the delayed-update effect. While Adam and AdamW introduce momentum and adaptive per-coordinate scaling, ZenFlow relies on channel-level gradient-energy concentration rather than optimizer-specific update dynamics. Our cross-optimizer measurements in \tref{tab:gradient_locality_cross_model} further show that this locality remains strong under AdamW.

\subsection{Determinism.}

{\system}'s decoupled updating process is fundamentally different from traditional nondeterministic asynchronous training methods~\cite{elastic_averaging_sgd_nips2015,asynchronous_sgd_iclr2018,ps_osdi2014,petuum_kdd15}, where update timing depends on variable factors such as system latency. {\system} ensures determinism: Given fixed seeds and configurations, the gradient selection, update and offloading schedules follow a \emph{predetermined order}, making runs fully reproducible.

In multi-GPU training, ZenFlow overlaps important and unimportant channel computation only within each GPU; inter-GPU synchronization still follows standard ZeRO semantics with a barrier at every step. Consequently, per-stage execution latency does not affect the deterministic ordering of parameter updates observed across workers. We use deterministic CUDA kernels and apply a fixed channel-ranking pipeline. Together, these guarantees ensure that ZenFlow produces bit-exact identical training results under the same configuration.


%% file: contents/implementation.tex
\section{Implementation}
\label{sec:implementation}
We build \system on DeepSpeed~\cite{deepspeed_kdd20} with $\sim$11K lines of Python code, without requiring any change to user training code. 
Our design extends DeepSpeed's ZeRO-Offload~\cite{ren2021zero_offload} and ZeRO-Infinity~\cite{zeroinfinity_sc21} backends by integrating importance-aware gradient selection and \textit{selective-optimizer} into the DeepSpeed runtime. We describe the key implementation components of \system below.

\phead{Fully Segmented Gradient Selection.} 
In fully sharded distributed training, model states such as gradients are partitioned across GPUs. A gradient tensor may be flattened and split across multiple GPUs, so a single input-channel tensor can span two devices. To support fine-grained importance tracking, we introduce a segment mapping table that uses \texttt{(segment\_id, offset)} tuples to index and manage gradient metadata. Each \texttt{segment\_id} corresponds to one selected important channel. This structure enables flexible tracking of important channels during training even when they are scattered across devices. Additionally, we customize PyTorch's \texttt{flatten} operations~\cite{pytorch_nips2019} to reorganize gradient storage from a row-major to a column-major layout, improving channel-wise access and manipulation.

\phead{Concurrent CPU-side Optimizer.}
To overlap GPU computation with CPU-side updates, \system initializes multiple CPU optimizer instances at setup. Each optimizer updates its assigned gradients—grouped into I/O-efficient buckets---as soon as data arrives from the GPU. 
Updates run in dedicated processes, with double-buffering and \texttt{shared\_memory} enabling zero-copy communication. Concurrency is carefully managed to minimize synchronization overhead.

\phead{Selective GPU-side Optimizer.} 
We extend PyTorch’s Adam and AdamW optimizers to support in-place updates using selected important gradients and enable fast swap-out and swap-in for extra optimizer state tensors.

%% file: contents/evaluation.tex
\section{Evaluation}
\label{sec:evaluation}

\subsection{Experimental Setup}
\label{subsec:experiment-setup}
\phead{Testbed.} 
The overall experimental environment is summarized in Table~\ref{tab:env}. 
Our experiments are conducted on two server configurations: an A100 testbed and an H100 testbed.
The A100 testbed consists of $4\times$ NVIDIA A100 GPUs (80GB each), fully interconnected via NVLink, and paired with 64 CPUs and 1TB of CPU memory. 
The H100 testbed uses $4\times$ NVIDIA H100 GPUs (80GB each) with NVLink, and paired with 64 CPUs and 2TB of CPU memory. Unless otherwise stated, experiments are conducted on the A100 testbed.

\phead{Models and Workloads.} 
We evaluate \system across a diverse set of LLMs: 
OPT-350M, Qwen2.5-\{0.5B, 1.5B, 3B\}, and Llama2-\{7B, 13B\}, spanning from 350M to 13B parameters. 
All models follow their original architecture and default hyperparameters. We fine-tune these models on the widely adopted GLUE benchmark~\cite{glue_2019} following the standard setup in prior work~\cite{jang2024smartinfinityfastlargelanguage}, and expand the evaluation to larger and more diverse models to reflect real-world usage. The selected models represent some of the most popular choices in the open-source community~\cite{hf_storage_arxiv25}, ensuring practical relevance.
Unless otherwise noted, we train each model for 3 epochs with a batch size of 8 and a learning rate of $1e\text{-}5$. 
We use the AdamW optimizer~\cite{adamw_2019} with a weight decay of 0.00 and apply a cosine learning rate schedule with 5\% warmup. These settings are aligned with prior studies~\cite{jang2024smartinfinityfastlargelanguage, galore_2024, lsp-offload_aaai25} to ensure fair, consistent comparison.

\begin{table}[t]
    \centering
    \caption{Experimental environments.}
    \renewcommand{\arraystretch}{1.0}
    \small
    \begin{adjustbox}{max width=0.95\columnwidth}
    \begin{tabular}{@{}l l l l@{}}
        \toprule
         & \textbf{ } & \textbf{A100 Testbed} & \textbf{H100 Testbed} \\
        \midrule
        \multirow{4}{*}{\textbf{HW}} 
            & GPU & NVIDIA A100 (80GB) ×4 & NVIDIA H100 (80GB) ×4 \\
            & CPU & \makecell[l]{AMD EPYC 7742 \\ 64C 128T (SMT Enabled)} 
                   & \makecell[l]{Intel Xeon Platinum 8462Y \\ 64C 128T (SMT Enabled)} \\
            & Memory & 32×32GB DDR4-3200 & 32×64GB DDR5-4800 \\
            & PCIe & PCIe 4.0 $\times$ 16 & PCIe 5.0 $\times$ 16 \\
        \midrule
        \multirow{4}{*}{\textbf{SW}} 
            & Python / PyTorch & \multicolumn{2}{c}{3.10 / 2.5.1} \\
            & CUDA / DeepSpeed & \multicolumn{2}{c}{11.8.0 / 0.16.2} \\
            & Model & \multicolumn{2}{c}{\makecell[c]{OPT-350M, Qwen2.5-\{0.5B, 1.5B, 3B\},\\ Llama-2-\{7B, 13B\}}} \\
        \bottomrule
    \end{tabular}
    \end{adjustbox}
    \label{tab:env}
\end{table}

\phead{Full-parameter fine-tuning baselines.} 
We compare \system (ZF) against SOTA offloading baselines: \textbf{ZeRO-Offload} (ZO)~\cite{ren2021zero_offload}, \textbf{ZeRO-Infinity}~\cite{zeroinfinity_sc21}, and \textbf{StrongHold} (SH)~\cite{stronghold_sc22}. 
We configure ZeRO-Offload with ZeRO Stage 2~\cite{rajbhandari2020zero} to maximize training throughput. For ZeRO-Infinity, we use default ZeRO Stage 3~\cite{zeroinfinity_sc21, rajbhandari2020zero} and disable NVMe offloading to avoid potential performance degradation caused by frequent SSD access during training. 
We implement StrongHold's layer-wise scheduling technique on top of DeepSpeed ZeRO-Offload because it is not open-sourced.

\phead{PEFT baseline.} We include GaLore~\cite{galore_2024} as a memory-saving PEFT baseline, using rank 128, batch size 8, and learning rate $4e\text{-}5$ with the default scale 0.25 to match ZenFlow's effective $1e\text{-}5$ update scale.
Notably, \system is orthogonal to these approaches and can be integrated with existing offloading strategies to further enhance scalability and performance---for example, by combining with gradient compression techniques or enabling deeper offloading with computational storage devices~\cite{jang2024smartinfinityfastlargelanguage}.

\begin{figure*}[!t]
    \centering
    \includegraphics[width=0.95\textwidth]{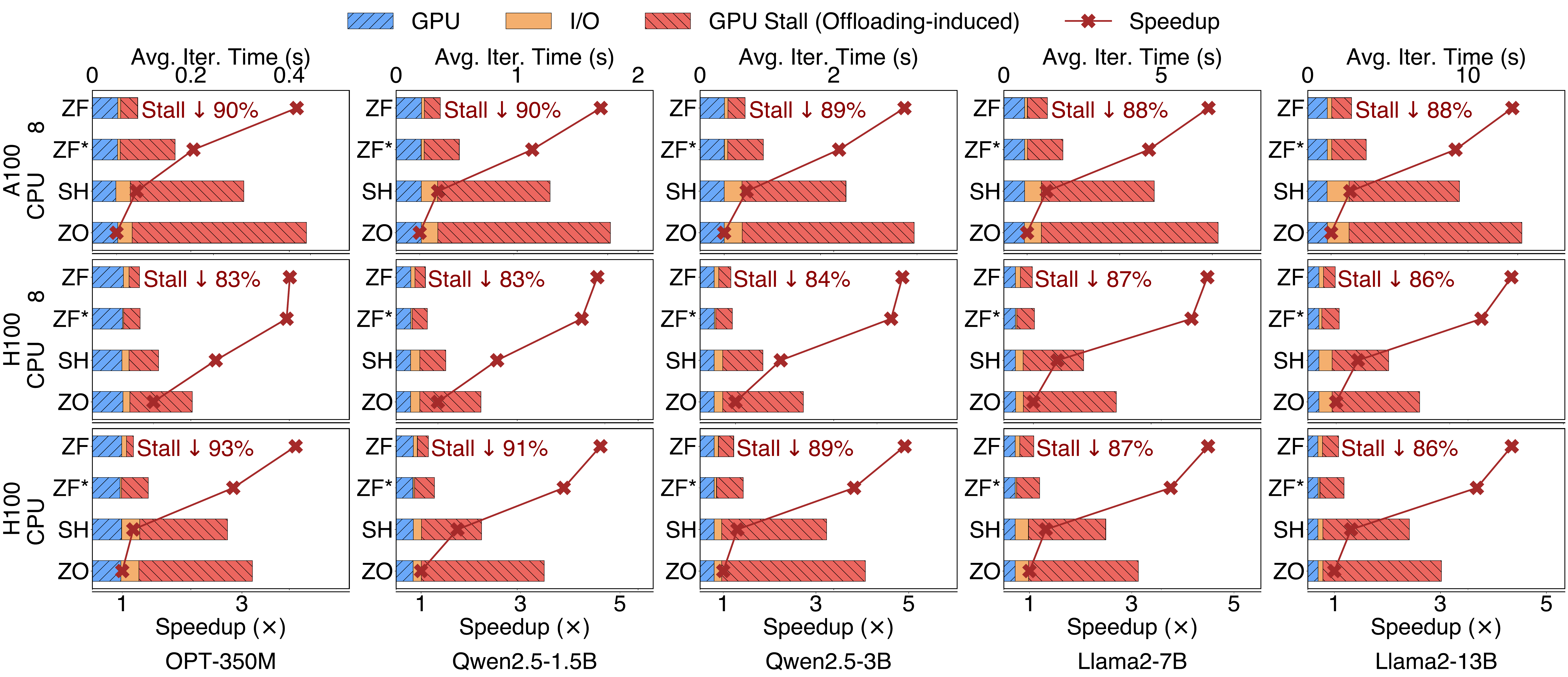}
    \caption{Training time breakdown and speedup across hardware configurations and models. Bars correspond to the upper X-axis. Lines (with markers) show relative speedup over ZeRO-Offload (ZO), referenced on the bottom X-axis. SH denotes StrongHold, ZF* denotes ZenFlow* (importance-aware updates without zero-stall pipelining), and ZF denotes ZenFlow (full system).} 
    \label{fig:iteration_time_breakdown}
    \Description{Fifteen bar-and-line charts compare training-time components and speedup for five models on A100 GPUs, H100 GPUs, and H100 GPUs with limited CPU resources. ZenFlow consistently removes most offloading-induced GPU stalls and provides the highest speedup.}
\end{figure*}

\phead{\system Variants.} We evaluate two variants of \system to isolate the contributions of its key design components. \textbf{\systemfull} represents the complete design of \system. It integrates two major optimizations: (1)~\textit{importance-aware offloading}, which selectively delays updates of unimportant parameters to reduce GPU stalls caused by CPU updates; 
and (2)~\textit{zero-stall pipeline}, 
which overlaps CPU-side optimizer updates with GPU computation and enables fast swap-out/in of \textit{selective-optimizer} states on the GPU to avoid memory footprint peaks (\sref{subsec:zero_bubble_pipeline}). \textbf{\systemasync} is a simplified variant that \emph{disables} the zero-stall pipeline while retaining importance-aware selective updates. This configuration isolates the performance benefit of pipelining. 

\phead{\system Hyperparameters.} 
\system introduces two additional hyperparameters: the update interval $S$ and the importance selection ratio \texttt{topk\_ratio} $k$.  
Unless otherwise specified, we set $S{=}4$, meaning less important parameters are updated once every 4 iterations and $k{=}0.01$, meaning top-1\% gradients are selected and kept updating on GPU (see \sref{subsec:zero_bubble_pipeline}). 
We provide a detailed analysis of these hyperparameter choices and their impact on accuracy and performance in \sref{subsec:sensitivity}.

\subsection{End-to-end Training Efficacy} 
\label{sec:eval_e2e_throughput}

\phead{Training Throughput.} 
We first evaluate the end-to-end training throughput of \system and compare it with state-of-the-art offloading baselines described above. We adopt representative fine-tuning configurations across different model scales: Qwen2.5-3B on 1 GPU with a batch size of 64, Llama2-7B on 2 GPUs with a batch size of 64, and Llama2-13B on 4 GPUs with a batch size of 48. Throughput is reported in TFLOPS using the DeepSpeed FLOPS  profiler~\cite{deepspeed_flops_profiler_2025}. 

\begin{figure}[t]
    \centering
    \begin{minipage}{0.75\textwidth}
    \begin{minipage}[t]{0.48\linewidth}
        \centering
        \includegraphics[width=\textwidth]{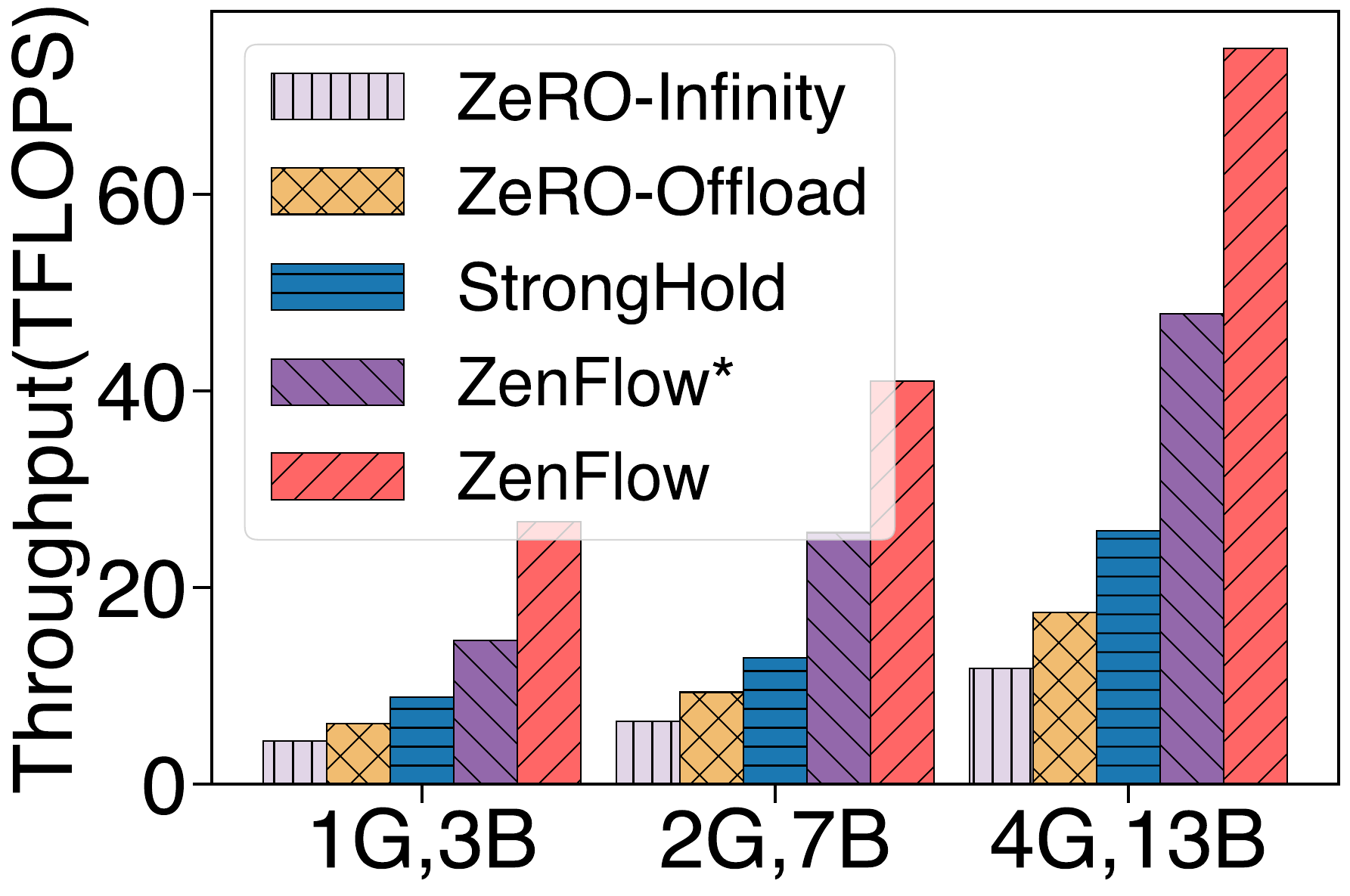}
        \caption{Throughput across different model and GPU count configurations.} 
        \label{fig:throughput}
        \Description{Bar chart comparing model throughput.}
    \end{minipage}\hfill
    \begin{minipage}[t]{0.48\linewidth}
        \centering
        \includegraphics[width=\textwidth]{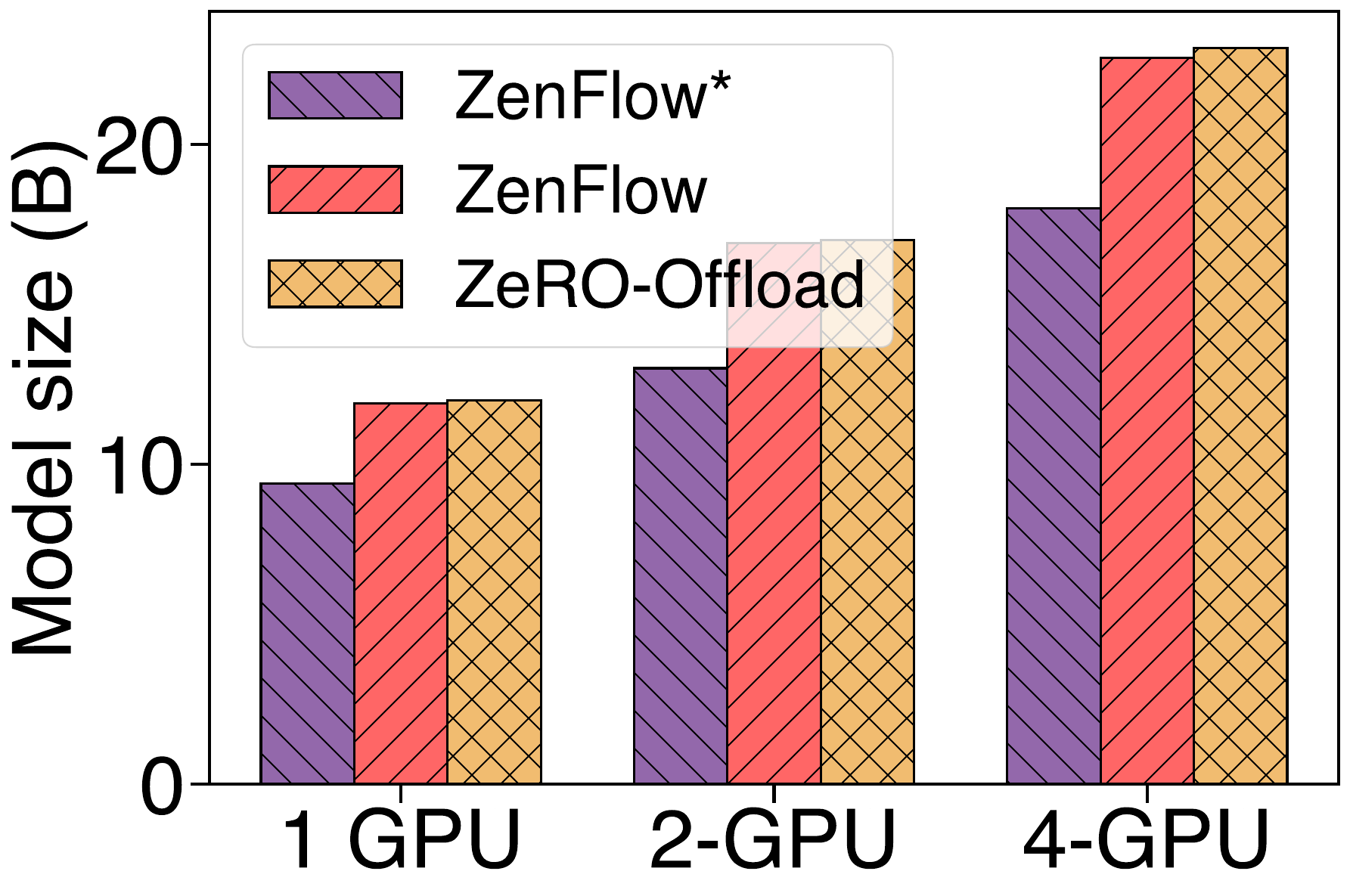}
        \caption{Maximum model size supported by each system vs. GPU count.}
        \label{fig:memory}
        \Description{Line chart showing model memory usage.}
    \end{minipage}
    \end{minipage}
\end{figure}

ZeRO-Infinity consistently exhibits the lowest throughput due to communication overhead incurred by ZeRO Stage 3~\cite{rajbhandari2020zero}. Therefore, in subsequent tests, we focus on baselines configured with ZeRO Stage 2, which achieve higher GPU utilization and more clearly highlight the impact of GPU stalls on end-to-end training. 
StrongHold improves over ZeRO-Offload by overlapping part of the GPU computation (i.e., backward pass) with CPU updates; 
however, the optimizer update phase remains dominant and limits its speedup. 
\systemasync introduces gradient decoupling based on importance, enabling notable performance gains. \system further improves efficiency by aggressively overlapping CPU updates within the accumulation phase. As shown in \fref{fig:throughput}, \system consistently outperforms all baselines across all configurations, achieving on average 4.3$\times$ speedup over ZeRO-Offload and  6.3$\times$ speedup over ZeRO-Infinity.

\phead{Model Scale.}  
We compare the maximum trainable model size under different systems. As shown in \fref{fig:memory}, both \system and ZeRO-Offload offload only optimizer states to ensure a fair comparison. 
\system achieves comparable model scalability across 1, 2, and 4 GPUs. 
\systemasync supports slightly smaller models due to additional GPU memory overhead incurred by maintaining a dedicated optimizer for important gradients.

\begin{figure}[ht]
    \centering
    \includegraphics[width=0.75\textwidth]{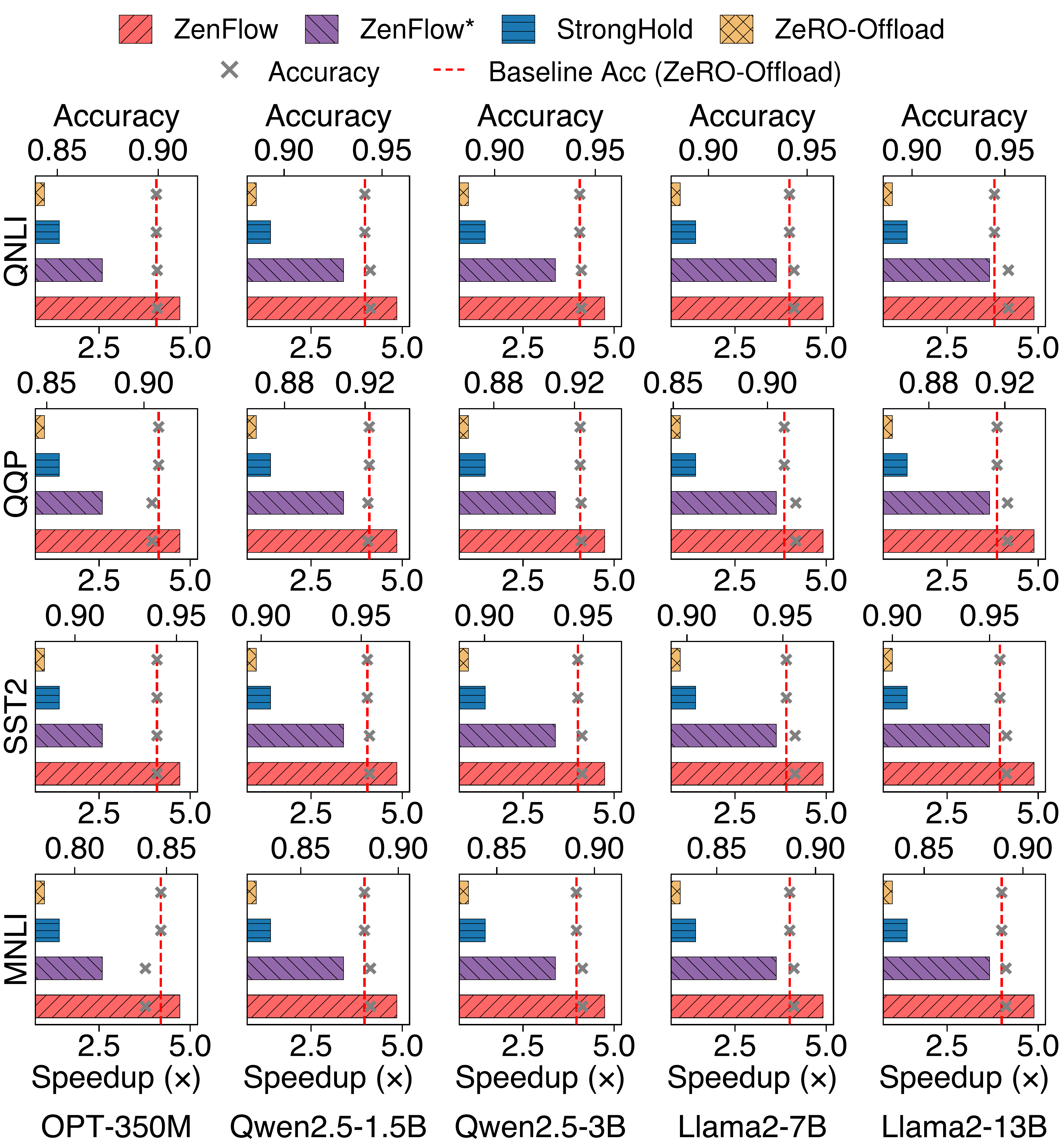}
    \caption{Accuracy vs. per-iteration speedup on GLUE tasks for various models and baselines. The red dashed line marks the baseline accuracy achieved by ZeRO-Offload.
Results close to the red line indicate accuracy comparable to the baseline while achieving significant speedup.} 
    \label{fig:accuracy}
    \Description{Accuracy vs. per-iteration speedup on GLUE tasks.} 
\end{figure}

\phead{Accuracy and Speedup.} 
Next, we evaluate \system and baseline methods on four representative GLUE tasks (MNLI, QNLI, QQP, and SST-2) across a range of model sizes and families. As shown in \fref{fig:accuracy}, \system achieves competitive or superior accuracy compared to existing offloading methods. 

Notably, for larger models such as Llama2-7B and Llama2-13B, \system consistently outperforms baselines. This improvement comes from our importance-aware design: By selectively prioritizing important channels during fine-tuning, \system preserves the essential learning capacity of the model with significant speedup gains. 

In some cases, such as with the smaller OPT-350M model, \system yields slightly lower accuracy due to the fixed update interval setting ($S=4$), which may be too coarse to capture rapid gradient changes during early training. This can be addressed via auto-tuning, as discussed in \sref{subsec:sensitivity}. 

\subsection{Time Breakdown and GPU Stall Analysis}
\label{sec:eval_time_breakdown}

We evaluate \system under three hardware configurations. The first setup reflects practical training environments with 4$\times$ A100 GPUs and full CPU parallelism. The second setup upgrades to 4$\times$ H100 GPUs with the same CPU capacity.
The third setting explores the impact of CPU under-provisioning---common in shared clusters where users share the same GPU node but are allocated only a small, exclusive portion of CPU resources (e.g., 8 cores per user).

As shown in \fref{fig:iteration_time_breakdown}, \system significantly reduces CPU-induced GPU stall time across all settings. By decoupling and overlapping CPU-side updates, \system consistently eliminates over 80\% of GPU stalls, leading to a 2.9$\times$--5$\times$ end-to-end speedup compared to ZeRO-Offload. For larger models (e.g., Llama2-7B and 13B), the benefits are more pronounced: the CPU-side overhead becomes a bottleneck due to heavy optimizer updates, even with highly parallelized, AVX-optimized \texttt{CPUAdam}.
\system effectively mitigates this bottleneck, reducing the CPU:GPU compute time ratio from over $12:1$ (in baseline runs on 7B / 13B models) to $1:1$ or lower. 

While {\system} introduces minor I/O overhead due to gradient gathering, these costs are effectively hidden by lightweight gradient selection and pipelined execution. The detailed breakdown of this overhead is presented in \sref{subsec:overhead_breakdown}.

\begin{figure}[t]
    \centering
    \includegraphics[width=0.75\textwidth]{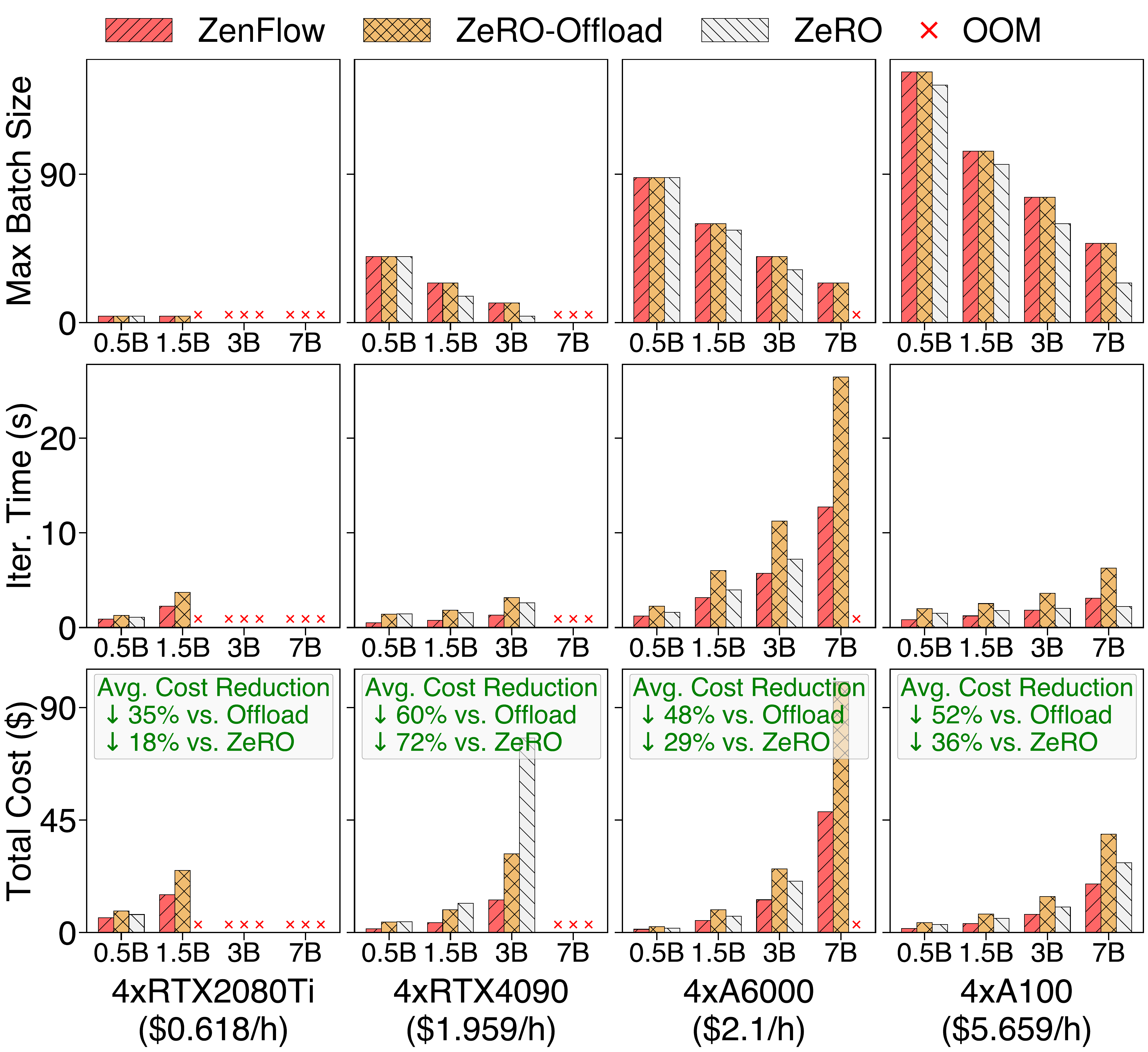}
    \caption{Cost comparison for fine-tuning Qwen (0.5B--7B) on Alpaca-52K with manually tuned maximum batch sizes.
We report the maximum feasible batch size (top), per-iteration time (middle), and total end-to-end training cost (bottom) across different GPU platforms.
ZenFlow enables lower training cost by (i) supporting larger batch sizes on cheaper hardware through offloading, and (ii) maintaining competitive iteration speed despite CPU-GPU heterogeneity.
As a result, ZenFlow achieves substantial reductions in end-to-end training cost compared to ZeRO-Offload and GPU-only ZeRO, while scaling to larger models without OOM.}
    \label{fig:use_case}
    \Description{Batch size, iteration time, and estimated cost for fine-tuning models of different sizes (0.5B-7B) on four hardware configurations (4$\times$ RTX2080Ti, 4$\times$ RTX4090, 4$\times$ A6000, and 4$\times$ A100), comparing ZenFlow, ZeRO-Offload, and ZeRO. Costs are calculated using the listed on-demand hourly rates. Red cross marks indicate out-of-memory configurations.} 
\end{figure}

\subsection{Practical Cost Savings}
\label{sec:cost_savings}

Using RunPod on-demand pricing~\cite{runpod-pricing}, we compare \system with ZeRO-Offload and GPU-only ZeRO, where GPU-only ZeRO is the full-GPU-memory baseline without CPU offloading. As shown in \fref{fig:use_case}, \system extends the model scale beyond what GPU ZeRO-only can support (e.g., 7B parameters on A6000), enabling fine-tuning of larger models without using more expensive hardware. In addition, \system improves efficiency: by leveraging offloading to support larger batch sizes and overlap updates, \system achieves higher end-to-end throughput than both ZeRO and ZeRO-Offload. This results in faster completion of the \texttt{Alpaca52K} fine-tuning task and lower overall cost. Even when the optimizer overhead is not fully amortized (e.g., 7B on A100), the larger batch sizes enabled by \system still reduce cost compared to traditional GPU training. Overall, \system reduces cost by 35--60\% compared to ZeRO-Offload and by 18--72\% compared to GPU-only ZeRO, demonstrating the practical benefits of importance-aware offloaded training.
Compared with GPU-only ZeRO, ZenFlow trades limited CPU offloading overhead for larger feasible batches and trainability when the model would otherwise run out of GPU memory.

\subsection{Convergence Validation}
\label{sec:eval_convergence}

\begin{figure}[ht]
    \centering
    \includegraphics[width=0.68\textwidth]{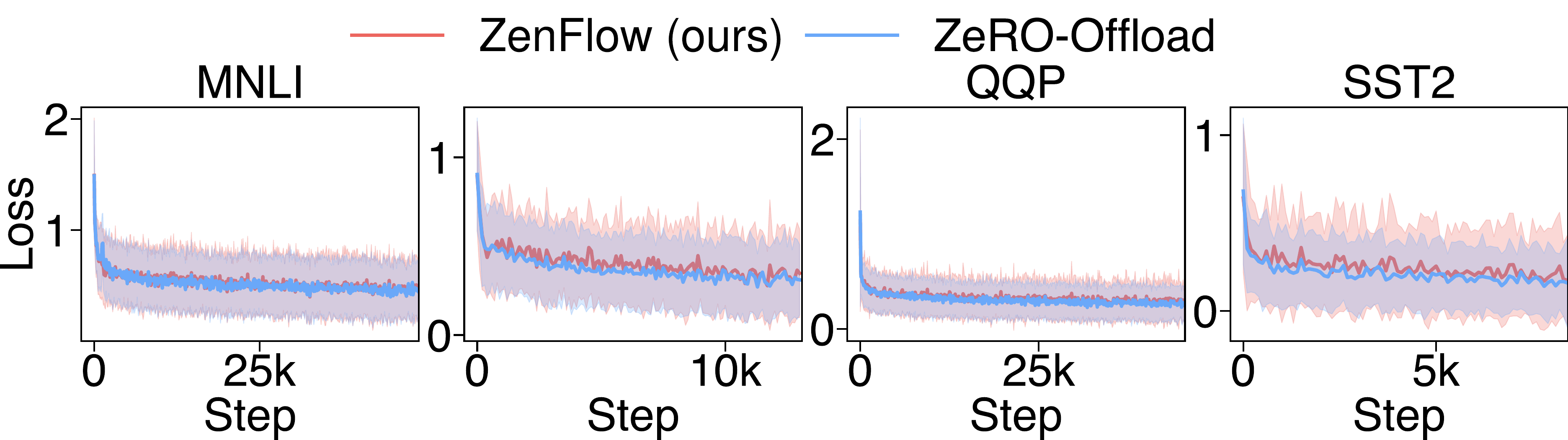}
    \caption{Convergence on GLUE with OPT-350M. \system's loss matches ZeRO-Offload across tasks.}
    \label{fig:convergence}
    \Description{Convergence comparison.}
\end{figure}

We evaluate the convergence behavior of \system compared to ZeRO-Offload on the GLUE benchmark using OPT-350M. As shown in \fref{fig:convergence}, we report the training loss on four representative GLUE tasks. 
Across all tasks, \system exhibits stable and competitive convergence. Its loss curves closely follow those of ZeRO-Offload, often with low variance (e.g., MNLI), converging at a similar rate in terms of iterations, but achieving much faster absolute convergence time due to reduced iteration latency. Results for other models follow similar trends and are omitted here due to space limit. 

\subsection{Gradient Locality Across Architectures}
\label{subsec:eval_gradient_locality}

As shown in \tref{tab:gradient_locality_cross_model}, strong channel locality appears across dense decoder-only, MoE, vision-language, and encoder-decoder models under both AdamW and SGD. The top-1\% channels capture 45.5--68.5\% of gradient $\ell_2^2$, i.e., 45.5--68.5$\times$ higher concentration than a uniform/random distribution, and their Jaccard similarity across steps is 0.40--0.81 versus a random baseline of $\approx$0.005.

\begin{table}[t]
\begingroup
\centering
\caption{Channel-level gradient $\ell_2^2$ concentration over 1,000 Alpaca-52K fine-tuning steps. Top-1\% coverage equals the concentration factor over uniform. Jaccard is computed between consecutive steps; random baseline $\approx$0.005.}
\label{tab:gradient_locality_cross_model}
\footnotesize
\setlength{\tabcolsep}{4pt}
\renewcommand{\arraystretch}{0.95}
\begin{adjustbox}{max width=0.88\columnwidth}
\begin{tabular}{lcccc}
\toprule
\textbf{Model} & \textbf{Opt.} & \textbf{Top-1\%} & \textbf{Top-10\%} & \textbf{Jaccard} \\
\midrule
Qwen2.5-7B & AdamW & 67.7\% & 81.6\% & 0.64 \\
Qwen1.5-MoE-A2.7B & AdamW & 54.8\% & 72.9\% & 0.54 \\
Qwen2-VL-7B-Instruct & AdamW & 62.9\% & 79.3\% & 0.72 \\
BART-large & AdamW & 68.5\% & 81.7\% & 0.40 \\
Qwen2.5-7B & SGD & 54.3\% & 71.4\% & 0.81 \\
Qwen1.5-MoE-A2.7B & SGD & 54.7\% & 72.3\% & 0.58 \\
Qwen2-VL-7B-Instruct & SGD & 45.5\% & 77.7\% & 0.53 \\
BART-large & SGD & 51.0\% & 73.5\% & 0.66 \\
\bottomrule
\end{tabular}
\end{adjustbox}
\endgroup
\end{table}

\subsection{Validation on MoE Architecture}
\label{subsec:eval_moe_validation}

We further validate ZenFlow on the Qwen1.5-MoE-A2.7B (14.3B parameters) MoE architecture. On QNLI/MNLI/QQP/SST-2, ZeRO-Offload/ZenFlow achieve 87.45/87.42, 84.93/85.45, 86.27/86.83, and 80.39/81.67, respectively. This comparable accuracy confirms that importance-aware channel selection also applies to sparse expert models.

\subsection{PEFT Comparison}
\label{subsec:eval_peft_comparison}

\tref{tab:peft_comparison_accuracy} compares GaLore and ZenFlow on GLUE accuracy across model scales on our A100 testbed. Because GaLore is not compatible with ZeRO-2 sharding and keeps optimizer-state memory, its memory footprint remains high; ZenFlow keeps comparable iteration time while reducing peak memory footprint (e.g., 7B from 53.2GB to 18.5GB) (\tref{tab:peft_comparison_system}).

\begin{table}[htbp]
\begingroup
\centering
\caption{PEFT comparison on GLUE accuracy (\%).}
\label{tab:peft_comparison_accuracy}
\footnotesize
\setlength{\tabcolsep}{3.5pt}
\renewcommand{\arraystretch}{0.95}
\begin{adjustbox}{max width=0.95\columnwidth}
\begin{tabular}{llccc}
\toprule
\textbf{Task} & \textbf{System} & \textbf{Qwen2.5-1.5B} & \textbf{Qwen2.5-3B} & \textbf{Llama2-7B} \\
\midrule
\multirow{2}{*}{QNLI}
& GaLore & 90.74 & 91.54 & 93.30 \\
& ZenFlow & 94.40 & 94.31 & 94.34 \\
\midrule
\multirow{2}{*}{QQP}
& GaLore & 86.82 & 87.53 & 85.10 \\
& ZenFlow & 92.14 & 92.33 & 90.93 \\
\midrule
\multirow{2}{*}{SST-2}
& GaLore & 94.95 & 95.76 & 96.79 \\
& ZenFlow & 95.41 & 95.61 & 95.64 \\
\midrule
\multirow{2}{*}{MNLI}
& GaLore & 84.00 & 84.20 & 87.30 \\
& ZenFlow & 88.71 & 89.52 & 88.73 \\
\bottomrule
\end{tabular}
\end{adjustbox}
\endgroup
\end{table}

\begin{table}[htbp]
\begingroup
\centering
\caption{PEFT comparison on system metrics on the A100 testbed.}
\label{tab:peft_comparison_system}
\footnotesize
\setlength{\tabcolsep}{3.5pt}
\renewcommand{\arraystretch}{0.95}
\begin{adjustbox}{max width=0.90\columnwidth}
\begin{tabular}{llccc}
\toprule
\textbf{System} & \textbf{Metric} & \textbf{Qwen2.5-1.5B} & \textbf{Qwen2.5-3B} & \textbf{Llama2-7B} \\
\midrule
ZenFlow & Memory footprint & 8.9GB & 13.2GB & 18.5GB \\
GaLore & Memory footprint & 15.5GB & 28.4GB & 53.2GB \\
\midrule
ZenFlow & Avg. iter. time & 372ms & 673ms & 1,322ms \\
GaLore & Avg. iter. time & 451ms & 660ms & 999ms \\
\bottomrule
\end{tabular}
\end{adjustbox}
\endgroup
\end{table}
\FloatBarrier
\subsection{Randomized Selection Ablation}
\label{subsec:eval_randomized_selection}

To isolate the effect of importance-aware selection, we compare ZenFlow with two randomized variants on Qwen2.5-3B: random selection keeps a random $k$ fraction of channels on GPU, while random discard drops the remaining gradients without CPU-side accumulation. As shown in \tref{tab:randomized_selection_ablation}, random selection consistently performs worse than ZenFlow, reducing average accuracy by 4.3 percentage points. This confirms that selecting high-importance channels, rather than merely retaining any channels under the same budget, is critical.

\begin{table}[htbp]
\begingroup
\centering
\caption{Randomized selection ablation on Qwen2.5-3B GLUE accuracy (\%).}
\label{tab:randomized_selection_ablation}
\footnotesize
\setlength{\tabcolsep}{4pt}
\renewcommand{\arraystretch}{0.95}
\begin{adjustbox}{max width=0.85\columnwidth}
\begin{tabular}{lccccc}
\toprule
\textbf{System} & \textbf{QNLI} & \textbf{QQP} & \textbf{SST-2} & \textbf{MNLI} & \textbf{Avg.} \\
\midrule
ZenFlow & 94.31 & 92.33 & 95.61 & 89.52 & 92.94 \\
Random discard & 93.65 & 91.70 & 93.69 & 88.17 & 91.80 \\
Random selection & 90.77 & 87.60 & 90.89 & 85.39 & 88.66 \\
\bottomrule
\end{tabular}
\end{adjustbox}
\endgroup
\end{table}

\FloatBarrier
\subsection{Broader Downstream Task Evaluation}
\label{subsec:eval_harder_tasks}

We further evaluate Qwen2.5-3B on MMLU and GSM8K, which measure knowledge and mathematical reasoning capabilities. We report standard zero-shot and few-shot accuracy to assess whether ZenFlow preserves general downstream capability after fine-tuning. As shown in \tref{tab:broader_downstream_tasks}, ZenFlow achieves comparable accuracy across shot settings, with only small task-dependent variations.

\begin{table}[t]
\begingroup
\centering
\caption{Broader downstream accuracy comparison (\%).}
\label{tab:broader_downstream_tasks}
\footnotesize
\setlength{\tabcolsep}{3.5pt}
\renewcommand{\arraystretch}{0.95}
\begin{adjustbox}{max width=0.88\columnwidth}
\begin{tabular}{llcccc}
\toprule
\textbf{Task} & \textbf{System} & \textbf{0-shot} & \textbf{5-shot} & \textbf{8-shot} & \textbf{10-shot} \\
\midrule
\multirow{2}{*}{MMLU}
& ZeRO-Offload & 50.59 & 53.71 & 54.88 & 59.38 \\
& ZenFlow & 50.59 & 54.10 & 54.68 & 59.08 \\
\midrule
\multirow{2}{*}{GSM8K}
& ZeRO-Offload & 2.34 & 63.67 & 68.36 & 69.92 \\
& ZenFlow & 2.34 & 63.37 & 70.80 & 71.88 \\
\bottomrule
\end{tabular}
\end{adjustbox}
\endgroup
\end{table}

\subsection{Sensitivity of Hyperparameters}
\label{subsec:sensitivity}

\begin{figure}[ht]
    \centering
    \includegraphics[width=0.75\textwidth]{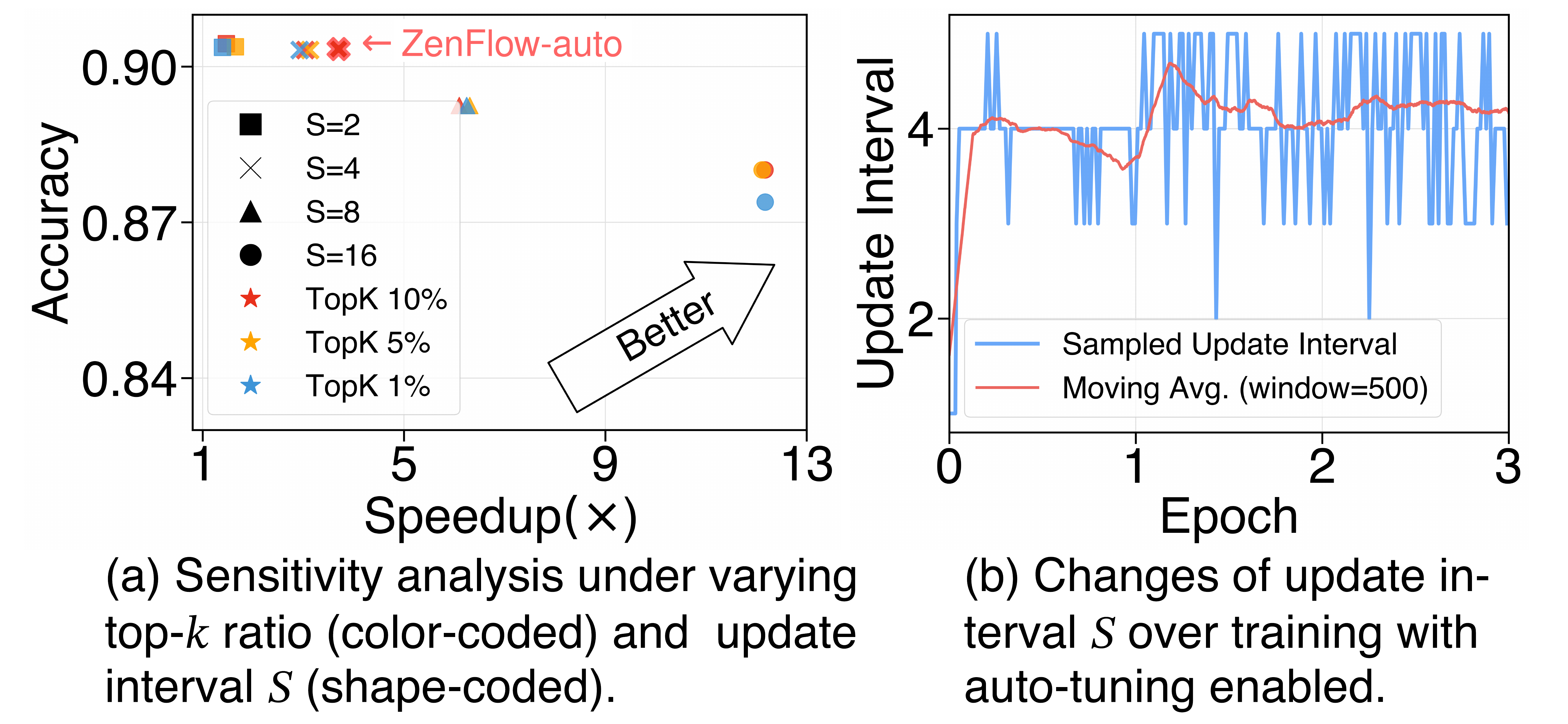}
    \caption{Sensitivity analysis of key hyperparameters.}
    \label{fig:sensitivity}
    \Description{Sensitivity of top-k ratio and update interval.}
\end{figure}

We investigate the impact of hyperparameters in \system, specifically the top-$k$ ratio and the CPU-side update interval. By assigning longer update intervals to less important gradients, \system further reduces GPU-side stalls. However, this selective staleness can slightly degrade accuracy when training iterations are limited---e.g., using $S{=}16$ results in a noticeable 0.02 drop in accuracy within just 3 epochs as shown in \fref{fig:sensitivity}(a) . This tradeoff can be mitigated with extended training iterations, as stale gradients gradually catch up due to \system's faster per-iteration execution.

From \fref{fig:sensitivity}(a) we observe that \system is generally robust to variations in the top-$k$ ratio in most cases, as \system preserves all gradients with bounded staleness. Increasing the update interval typically improves speedup. However, using an overly aggressive update interval (e.g., $S{=}16$), a smaller ratio may degrade accuracy due to increased staleness penalty (e.g., from 0.881 to 0.873 with 1\% top-$k$). In general, we find that a 1\% ratio with $S{=}4$ strikes a good balance between accuracy and efficiency (see \sref{subsec:problem_insight}).

\phead{Warmup.} Warmup lets ZenFlow perform channel selection after early gradients stabilize. On Qwen2.5-7B, top-1\% channel coverage rises from 63.8\% early in training to 67.7\% later. We vary the warmup ratio on Qwen2.5-3B; as shown in \tref{tab:warmup_sensitivity}, 5\% warmup provides a good tradeoff, preserving strong speedup while avoiding the QNLI accuracy drop observed without warmup.

\begin{table}[t]
\begingroup
\centering
\caption{Warmup-ratio sensitivity on GLUE accuracy (\%).}
\label{tab:warmup_sensitivity}
\footnotesize
\setlength{\tabcolsep}{3pt}
\renewcommand{\arraystretch}{0.95}
\begin{adjustbox}{max width=0.90\columnwidth}
\begin{tabular}{lcccccc}
\toprule
\textbf{Warmup} & \textbf{QNLI} & \textbf{QQP} & \textbf{SST-2} & \textbf{MNLI} & \textbf{Avg.} & \textbf{Speedup} \\
\midrule
0\% & 93.84 & 92.27 & 95.89 & 89.11 & 92.78 & 4.97$\times$ \\
5\% & 94.31 & 92.33 & 95.61 & 89.52 & 92.94 & 4.69$\times$ \\
10\% & 94.32 & 92.18 & 95.61 & 89.78 & 92.97 & 4.14$\times$ \\
\bottomrule
\end{tabular}
\end{adjustbox}
\endgroup
\end{table}

To empirically balance accuracy and speedup, we analyze how quickly less important gradients accumulate to match the significance of important ones. For instance, a low-importance gradient starting at 0.1 may grow to 0.5 within five iterations, approaching the magnitude of high-importance gradients. This  observation informs our auto-update mechanism shown in \fref{fig:sensitivity}(b): early in training, the update interval is kept short (e.g., 1-2), ensuring responsiveness, while later it is adaptively relaxed to 4–5 as training stabilizes. This dynamic configuration delivers higher speedups in later stages without compromising final accuracy, outperforming fixed configurations such as $S{=}4$.  

\subsection{Communication Overhead Breakdown}
\label{subsec:overhead_breakdown} 

\system's lightweight gradient gathering dramatically reduces communication volume and incurs minimal runtime overhead. As shown in \fref{fig:communication_overhead}, it achieves over 6,000$\times$ reduction in communication volume and adds less than 0.2s overhead per iteration—substantially lower than full gradient gathering, even on large models like Llama2-13B.

\subsection{Applicability under TP and PP}
\label{subsec:applicability_tp_pp}
\begin{figure}[ht]
    \centering
    \begin{minipage}{0.75\textwidth}
    \begin{minipage}[t]{0.48\linewidth}
        \centering
        \includegraphics[width=\textwidth]{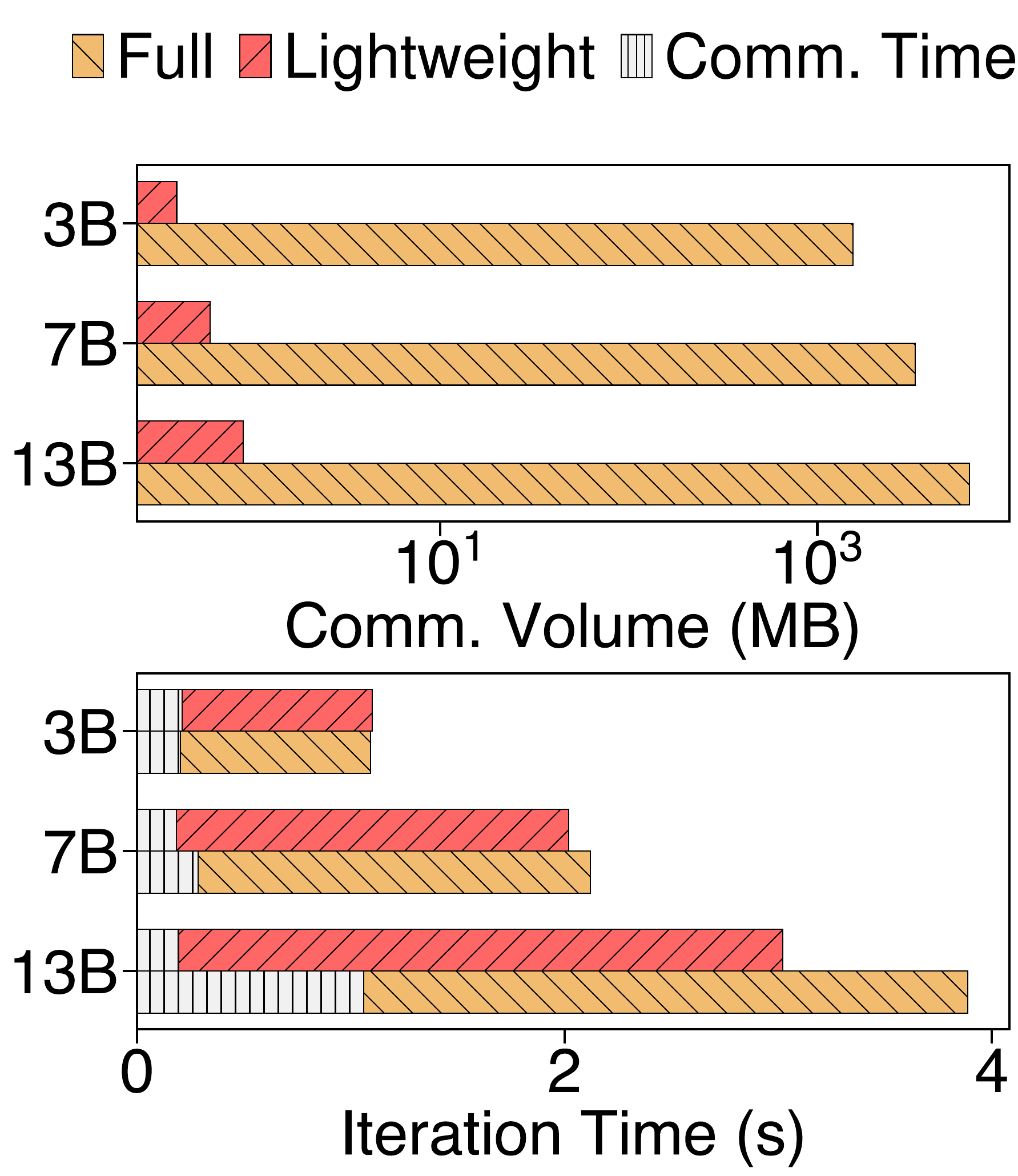}
        \caption{Gradient gathering communication overhead breakdown.} 
        \label{fig:communication_overhead}
        \Description{Bar chart comparing communication volume.}
    \end{minipage}\hfill
    \begin{minipage}[t]{0.48\linewidth}
        \centering
        \includegraphics[width=\textwidth]{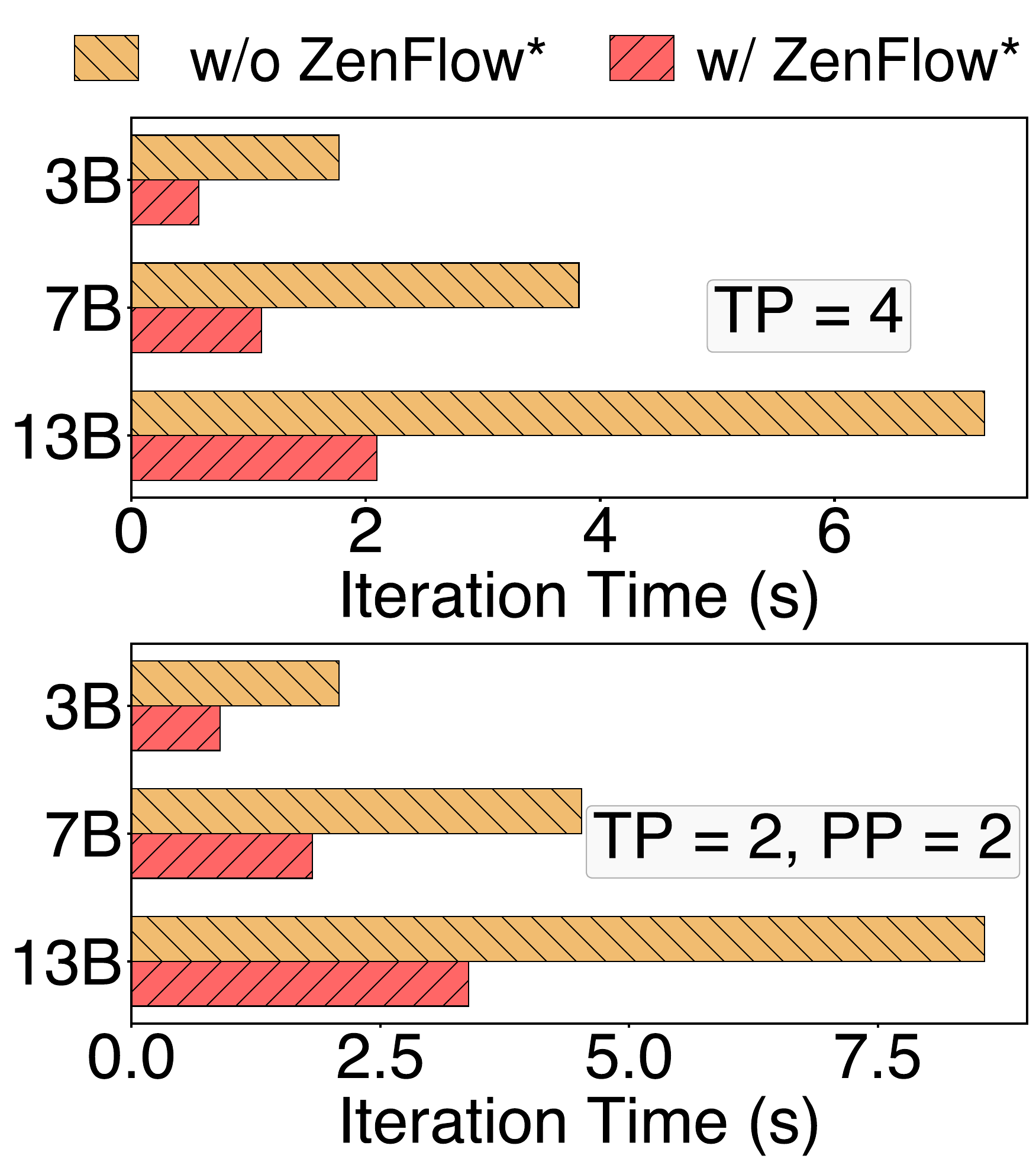}
        \caption{Iteration time with and without ZenFlow* under TP and TP+PP.} 
        \label{fig:tp_pp}
        \Description{Throughput and parallel efficiency.}
    \end{minipage}
    \end{minipage}
\end{figure}

We evaluate \system with tensor parallelism (TP) and pipeline parallelism (PP) by enabling ZenFlow's importance-aware offloading workflow in Megatron-LM~\cite{shoeybi2019megatron}. \fref{fig:tp_pp} shows \systemasync reduces iteration time under both TP=4 and TP=2+PP=2 on $4\times$ A100, even for 13B models. By reducing unnecessary compute and I/O overhead for unimportant parameters, \system’s gains are parallelism-agnostic and hold beyond ZeRO.

%% file: contents/related_work.tex
\section{Related Work}
\label{sec:related_work}

\phead{Tensor Offloading for Training.} 
Prior work~\cite{ren2021zero_offload,zeroinfinity_sc21,stronghold_sc22,NEURIPS2021_Efie_offloadDNN} offloads tensors to CPU memory to scale training but largely overlooks CPU--GPU pipeline interactions, leaving stalls from CPU latency and PCIe transfers~\cite{FlexScience24_BreakMemWall}. These systems make large models trainable under GPU memory constraints, yet generally update all parameters uniformly despite the hardware imbalance. StrongHold~\cite{stronghold_sc22} overlaps layer-wise CPU and GPU computation, but the performance gap between the devices still exposes CPU-induced stalls; overlap alone therefore cannot fully hide slow CPU updates. Storage-focused systems such as SwapAdvisor~\cite{swapadvisor_asplos2020} and G10~\cite{g10_micro2023} optimize placement across GPU, CPU, and SSD while treating the CPU primarily as storage rather than explicitly coordinating its computation with GPU updates. 
Recent systems exploit high-bandwidth CPU--GPU superchips for efficient offloading~~\cite{superoffload_asplos2026,c2cserve_arxiv26}, 
but rely on specialized interconnects such as NVLink-C2C that are unavailable on widely deployed A100 and H100 GPUs.
In contrast, \system is importance- and hardware-aware, decoupling important and non-important updates to minimize stalls.

\phead{Gradient Sparsity and Compression.} 
Prior studies show that gradients exhibit sparsity and can be filtered using threshold-based selection~\cite{strom2015scalable,dryden2016communication,aji2017sparse,maskllm_nips2024,huang2025castcontinuousdifferentiablesemistructured,lin2017dgc}; transmitting only top-$k$ or above-threshold gradients can preserve training quality while reducing communication. In offloading settings, this observation motivates prioritizing useful gradients instead of moving every gradient with the same urgency. Smart-Infinity~\cite{jang2024smartinfinityfastlargelanguage} and LSP-Offload~\cite{lsp-offload_aaai25} reduce offloading I/O through lossy compression, dropping small gradients or using learned projections at the cost of gradient fidelity. \system instead retains the remaining gradients through CPU-side accumulation while scheduling important updates on the GPU. Compression remains orthogonal to \system and can further optimize its scheduling pipeline.

\phead{Asynchronous Training.} Asynchrony can accelerate distributed training~\cite{elastic_averaging_sgd_nips2015,asynchronous_sgd_iclr2018,ps_osdi2014,petuum_kdd15,stellaris_sc24,dlrover-rm_vldb25,fedat_sc21}, but non-deterministic execution introduces stale gradients that can degrade convergence and reproducibility~\cite{staleness_arxiv2018,asynchronous_sgd_nips2015,zhang2015staleness,convergence_PMLR2016}. In typical schemes, workers update shared state according to unpredictable completion times, making the resulting update sequence difficult to reproduce. Prior work therefore bounds staleness~\cite{stellaris_sc24,asynchronous_sgd_nips2015,fedat_sc21,ssp_atc14}. \system instead uses bounded asynchronous execution specifically to hide CPU-side latency while retaining a predetermined, repeatable update order.

\phead{Parameter-Efficient vs. Full-Parameter Fine-Tuning.} Parameter-Efficient Fine-Tuning (PEFT) methods (e.g., adapters~\cite{sparseadapter_arxiv2022}, prefix-tuning~\cite{prefix_tuning_arxiv2021}, LoRA~\cite{lora_iclr2022}, Q-LoRA~\cite{qlora_nips2023}, mLoRA~\cite{m-lora_vldb25}, GaLore~\cite{galore_2024}) reduce fine-tuning cost by updating only a small subset of parameters, significantly lowering memory and optimizer overhead.
However, this efficiency comes at a cost: the constrained parameter space can bottleneck the model's ability to learn from diverse datasets or sustain long-term training~\cite{revisiting_acl2022, parameter_acl2023, parameter_2025arxiv, dai2024discovering, patel2025guide}.
Unlike PEFT, \system targets full-parameter fine-tuning rather than restricting the trainable parameter space, preserving full-weight flexibility while improving system performance. The two approaches address different bottlenecks: \system is complementary to PEFT and can be integrated with low-rank methods.

%% file: contents/conclusion.tex
\section{Conclusion}
\label{sec:conclusion}


{\system} decouples fast GPU and slow CPU updates to reduce I/O and GPU stalls in LLM fine-tuning. It overlaps in-place GPU updates of important gradients with CPU accumulation of the remainder. By exploiting gradients' spatial and temporal locality for scalable importance estimation, {\system} accelerates fine-tuning without global synchronization while preserving accuracy and reproducibility. {\system} has been integrated into DeepSpeed.

%% file: cite.bib
@article{zhao2023pytorch_fsdp,
  title={Pytorch fsdp: experiences on scaling fully sharded data parallel},
  author={Zhao, Yanli and Gu, Andrew and Varma, Rohan and Luo, Liang and Huang, Chien-Chin and Xu, Min and Wright, Less and Shojanazeri, Hamid and Ott, Myle and Shleifer, Sam and others},
  journal={arXiv preprint arXiv:2304.11277},
  year={2023}
}

@article{shoeybi2019megatron,
  title={Megatron-lm: Training multi-billion parameter language models using model parallelism},
  author={Shoeybi, Mohammad and Patwary, Mostofa and Puri, Raul and LeGresley, Patrick and Casper, Jared and Catanzaro, Bryan},
  journal={arXiv preprint arXiv:1909.08053},
  year={2019}
}

@inproceedings{ssp_atc14,
author = {Cui, Henggang and Cipar, James and Ho, Qirong and Kim, Jin Kyu and Lee, Seunghak and Kumar, Abhimanu and Wei, Jinliang and Dai, Wei and Ganger, Gregory R. and Gibbons, Phillip B. and Gibson, Garth A. and Xing, Eric P.},
title = {Exploiting bounded staleness to speed up big data analytics},
year = {2014},
isbn = {9781931971102},
publisher = {USENIX Association},
address = {USA},
booktitle = {Proceedings of the 2014 USENIX Conference on USENIX Annual Technical Conference},
pages = {37–48},
numpages = {12},
location = {Philadelphia, PA},
series = {USENIX ATC'14}
}

@misc{c2cserve_arxiv26,
      title={C2CServe: Leveraging NVLink-C2C for Elastic Serverless LLM Serving on MIG}, 
      author={Shutian Luo and Ali Zafar Sadiq and Rui Yang and Mingye Zhang and Haiying Shen and Wei Wang and Yue Cheng},
      year={2026},
      eprint={2605.19481},
      archivePrefix={arXiv},
      primaryClass={cs.OS},
      url={https://arxiv.org/abs/2605.19481}, 
}

@misc{efficient_llm_survey_arxiv24,
      title={Beyond Efficiency: A Systematic Survey of Resource-Efficient Large Language Models}, 
      author={Guangji Bai and Zheng Chai and Chen Ling and Shiyu Wang and Jiaying Lu and Nan Zhang and Tingwei Shi and Ziyang Yu and Mengdan Zhu and Yifei Zhang and Xinyuan Song and Carl Yang and Yue Cheng and Liang Zhao},
      year={2024},
      eprint={2401.00625},
      archivePrefix={arXiv},
      primaryClass={cs.LG},
      url={https://arxiv.org/abs/2401.00625}, 
}

@inproceedings{rajbhandari2020zero,
  title={Zero: Memory optimizations toward training trillion parameter models},
  author={Rajbhandari, Samyam and Rasley, Jeff and Ruwase, Olatunji and He, Yuxiong},
  booktitle={SC20: International Conference for High Performance Computing, Networking, Storage and Analysis},
  pages={1--16},
  year={2020},
  organization={IEEE}
}

@inproceedings{ren2021zero_offload,
  title={$\{$Zero-offload$\}$: Democratizing $\{$billion-scale$\}$ model training},
  author={Ren, Jie and Rajbhandari, Samyam and Aminabadi, Reza Yazdani and Ruwase, Olatunji and Yang, Shuangyan and Zhang, Minjia and Li, Dong and He, Yuxiong},
  booktitle={2021 USENIX Annual Technical Conference (USENIX ATC 21)},
  pages={551--564},
  year={2021}
}

@article{strom2015scalable,
  title={Scalable distributed DNN training using commodity GPU cloud computing},
  author={Str{\"o}m, Nikko},
  year={2015}
}

@inproceedings{dryden2016communication,
  title={Communication quantization for data-parallel training of deep neural networks},
  author={Dryden, Nikoli and Moon, Tim and Jacobs, Sam Ade and Van Essen, Brian},
  booktitle={2016 2nd Workshop on machine learning in hpc environments (MLHPC)},
  pages={1--8},
  year={2016},
  organization={IEEE}
}

@misc{jang2024smartinfinityfastlargelanguage,
      title={Smart-Infinity: Fast Large Language Model Training using Near-Storage Processing on a Real System}, 
      author={Hongsun Jang and Jaeyong Song and Jaewon Jung and Jaeyoung Park and Youngsok Kim and Jinho Lee},
      year={2024},
      eprint={2403.06664},
      archivePrefix={arXiv},
      primaryClass={cs.AR},
      url={https://arxiv.org/abs/2403.06664}, 
}

@article{jaszczur2021sparse,
  title={Sparse is enough in scaling transformers},
  author={Jaszczur, Sebastian and Chowdhery, Aakanksha and Mohiuddin, Afroz and Kaiser, Lukasz and Gajewski, Wojciech and Michalewski, Henryk and Kanerva, Jonni},
  journal={Advances in Neural Information Processing Systems},
  volume={34},
  pages={9895--9907},
  year={2021}
}

@inproceedings{shen2020powernorm,
  title={Powernorm: Rethinking batch normalization in transformers},
  author={Shen, Sheng and Yao, Zhewei and Gholami, Amir and Mahoney, Michael and Keutzer, Kurt},
  booktitle={International conference on machine learning},
  pages={8741--8751},
  year={2020},
  organization={PMLR}
}

@article{koloskova2022sharper,
  title={Sharper convergence guarantees for asynchronous SGD for distributed and federated learning},
  author={Koloskova, Anastasiia and Stich, Sebastian U and Jaggi, Martin},
  journal={Advances in Neural Information Processing Systems},
  volume={35},
  pages={17202--17215},
  year={2022}
}

@article{zhang2015staleness,
  title={Staleness-aware async-sgd for distributed deep learning},
  author={Zhang, Wei and Gupta, Suyog and Lian, Xiangru and Liu, Ji},
  journal={arXiv preprint arXiv:1511.05950},
  year={2015}
}

@article{luo2025multi,
  title={A Multi-Power Law for Loss Curve Prediction Across Learning Rate Schedules},
  author={Luo, Kairong and Wen, Haodong and Hu, Shengding and Sun, Zhenbo and Liu, Zhiyuan and Sun, Maosong and Lyu, Kaifeng and Chen, Wenguang},
  journal={arXiv preprint arXiv:2503.12811},
  year={2025}
}

@article{hoffmann2022training,
  title={Training compute-optimal large language models},
  author={Hoffmann, Jordan and Borgeaud, Sebastian and Mensch, Arthur and Buchatskaya, Elena and Cai, Trevor and Rutherford, Eliza and Casas, Diego de Las and Hendricks, Lisa Anne and Welbl, Johannes and Clark, Aidan and others},
  journal={arXiv preprint arXiv:2203.15556},
  year={2022}
}

@article{kaplan2020scaling,
  title={Scaling laws for neural language models},
  author={Kaplan, Jared and McCandlish, Sam and Henighan, Tom and Brown, Tom B and Chess, Benjamin and Child, Rewon and Gray, Scott and Radford, Alec and Wu, Jeffrey and Amodei, Dario},
  journal={arXiv preprint arXiv:2001.08361},
  year={2020}
}

@inproceedings{shazeer2018adafactor,
  title={Adafactor: Adaptive learning rates with sublinear memory cost},
  author={Shazeer, Noam and Stern, Mitchell},
  booktitle={Proceedings of the 35th International Conference on Machine Learning},
  year={2018}
}

@article{loshchilov2017decoupled,
  title={Decoupled weight decay regularization},
  author={Loshchilov, Ilya and Hutter, Frank},
  journal={arXiv preprint arXiv:1711.05101},
  year={2017}
}

@inproceedings{stronghold_sc22,
  title={Stronghold: fast and affordable billion-scale deep learning model training},
  author={Sun, Xiaoyang and Wang, Wei and Qiu, Shenghao and Yang, Renyu and Huang, Songfang and Xu, Jie and Wang, Zheng},
  booktitle={SC22: International Conference for High Performance Computing, Networking, Storage and Analysis},
  pages={1--17},
  year={2022},
  organization={IEEE}
}

@inproceedings{NEURIPS2021_Efie_offloadDNN,
 author = {Beaumont, Olivier and Eyraud-Dubois, Lionel and Shilova, Alena},
 booktitle = {Advances in Neural Information Processing Systems},
 editor = {M. Ranzato and A. Beygelzimer and Y. Dauphin and P.S. Liang and J. Wortman Vaughan},
 pages = {23844--23857},
 publisher = {Curran Associates, Inc.},
 title = {Efficient Combination of Rematerialization and Offloading for Training DNNs},
 url = {https://proceedings.neurips.cc/paper_files/paper/2021/file/c8461bf13fca8a2b9912ab2eb1668e4b-Paper.pdf},
 volume = {34},
 year = {2021}
}

@inproceedings{zeroinfinity_sc21,
author = {Rajbhandari, Samyam and Ruwase, Olatunji and Rasley, Jeff and Smith, Shaden and He, Yuxiong},
title = {ZeRO-infinity: breaking the GPU memory wall for extreme scale deep learning},
year = {2021},
isbn = {9781450384421},
publisher = {Association for Computing Machinery},
address = {New York, NY, USA},
url = {https://doi.org/10.1145/3458817.3476205},
doi = {10.1145/3458817.3476205},
booktitle = {Proceedings of the International Conference for High Performance Computing, Networking, Storage and Analysis},
articleno = {59},
numpages = {14},
location = {St. Louis, Missouri},
series = {SC '21}
}

@inproceedings{FlexScience24_BreakMemWall,
author = {Maurya, Avinash and Ye, Jie and Rafique, M. Mustafa and Cappello, Franck and Nicolae, Bogdan},
title = {Breaking the Memory Wall: A Study of I/O Patterns and GPU Memory Utilization for Hybrid CPU-GPU Offloaded Optimizers},
year = {2024},
isbn = {9798400706424},
publisher = {Association for Computing Machinery},
address = {New York, NY, USA},
url = {https://doi.org/10.1145/3659995.3660038},
doi = {10.1145/3659995.3660038},
booktitle = {Proceedings of the 14th Workshop on AI and Scientific Computing at Scale Using Flexible Computing Infrastructures},
pages = {9–16},
numpages = {8},
location = {Pisa, Italy},
series = {FlexScience'24}
}

@misc{hf_storage_arxiv25,
      title={Towards Efficient LLM Storage Reduction via Tensor Deduplication and Delta Compression}, 
      author={Zirui Wang and Tingfeng Lan and Zhaoyuan Su and Juncheng Yang and Yue Cheng},
      year={2025},
      eprint={2505.06252},
      archivePrefix={arXiv},
      primaryClass={cs.DB},
      url={https://arxiv.org/abs/2505.06252}, 
}

@inproceedings{gpt3_nips20,
author = {Brown, Tom B. and Mann, Benjamin and Ryder, Nick and Subbiah, Melanie and Kaplan, Jared and Dhariwal, Prafulla and Neelakantan, Arvind and Shyam, Pranav and Sastry, Girish and Askell, Amanda and Agarwal, Sandhini and Herbert-Voss, Ariel and Krueger, Gretchen and Henighan, Tom and Child, Rewon and Ramesh, Aditya and Ziegler, Daniel M. and Wu, Jeffrey and Winter, Clemens and Hesse, Christopher and Chen, Mark and Sigler, Eric and Litwin, Mateusz and Gray, Scott and Chess, Benjamin and Clark, Jack and Berner, Christopher and McCandlish, Sam and Radford, Alec and Sutskever, Ilya and Amodei, Dario},
title = {Language models are few-shot learners},
year = {2020},
isbn = {9781713829546},
publisher = {Curran Associates Inc.},
address = {Red Hook, NY, USA},
booktitle = {Proceedings of the 34th International Conference on Neural Information Processing Systems},
articleno = {159},
numpages = {25},
location = {Vancouver, BC, Canada},
series = {NIPS '20}
}

@article{gpt2_2019,
  title={Language Models are Unsupervised Multitask Learners},
  author={Radford, Alec and Wu, Jeff and Child, Rewon and Luan, David and Amodei, Dario and Sutskever, Ilya},
  year={2019}
}

@misc{llama_2023,
      title={LLaMA: Open and Efficient Foundation Language Models}, 
      author={Hugo Touvron and Thibaut Lavril and Gautier Izacard and Xavier Martinet and Marie-Anne Lachaux and Timothée Lacroix and Baptiste Rozière and Naman Goyal and Eric Hambro and Faisal Azhar and Aurelien Rodriguez and Armand Joulin and Edouard Grave and Guillaume Lample},
      year={2023},
      eprint={2302.13971},
      archivePrefix={arXiv},
      primaryClass={cs.CL},
      url={https://arxiv.org/abs/2302.13971}, 
}

@misc{llama2_2023,
      title={Llama 2: Open Foundation and Fine-Tuned Chat Models}, 
      author={Hugo Touvron and Louis Martin and Kevin Stone and Peter Albert and Amjad Almahairi and Yasmine Babaei and Nikolay Bashlykov and Soumya Batra and Prajjwal Bhargava and Shruti Bhosale and Dan Bikel and Lukas Blecher and Cristian Canton Ferrer and Moya Chen and Guillem Cucurull and David Esiobu and Jude Fernandes and Jeremy Fu and Wenyin Fu and Brian Fuller and Cynthia Gao and Vedanuj Goswami and Naman Goyal and Anthony Hartshorn and Saghar Hosseini and Rui Hou and Hakan Inan and Marcin Kardas and Viktor Kerkez and Madian Khabsa and Isabel Kloumann and Artem Korenev and Punit Singh Koura and Marie-Anne Lachaux and Thibaut Lavril and Jenya Lee and Diana Liskovich and Yinghai Lu and Yuning Mao and Xavier Martinet and Todor Mihaylov and Pushkar Mishra and Igor Molybog and Yixin Nie and Andrew Poulton and Jeremy Reizenstein and Rashi Rungta and Kalyan Saladi and Alan Schelten and Ruan Silva and Eric Michael Smith and Ranjan Subramanian and Xiaoqing Ellen Tan and Binh Tang and Ross Taylor and Adina Williams and Jian Xiang Kuan and Puxin Xu and Zheng Yan and Iliyan Zarov and Yuchen Zhang and Angela Fan and Melanie Kambadur and Sharan Narang and Aurelien Rodriguez and Robert Stojnic and Sergey Edunov and Thomas Scialom},
      year={2023},
      eprint={2307.09288},
      archivePrefix={arXiv},
      primaryClass={cs.CL},
      url={https://arxiv.org/abs/2307.09288}, 
}

@misc{adam_2017,
      title={Adam: A Method for Stochastic Optimization}, 
      author={Diederik P. Kingma and Jimmy Ba},
      year={2017},
      eprint={1412.6980},
      archivePrefix={arXiv},
      primaryClass={cs.LG},
      url={https://arxiv.org/abs/1412.6980}, 
}

@inproceedings{gpipe_nips2019,
 author = {Huang, Yanping and Cheng, Youlong and Bapna, Ankur and Firat, Orhan and Chen, Dehao and Chen, Mia and Lee, HyoukJoong and Ngiam, Jiquan and Le, Quoc V and Wu, Yonghui and Chen, zhifeng},
 booktitle = {Advances in Neural Information Processing Systems},
 editor = {H. Wallach and H. Larochelle and A. Beygelzimer and F. d\textquotesingle Alch\'{e}-Buc and E. Fox and R. Garnett},
 pages = {},
 publisher = {Curran Associates, Inc.},
 title = {GPipe: Efficient Training of Giant Neural Networks using Pipeline Parallelism},
 url = {https://proceedings.neurips.cc/paper_files/paper/2019/file/093f65e080a295f8076b1c5722a46aa2-Paper.pdf},
 volume = {32},
 year = {2019}
}

@inproceedings{pipedream_sosp2019,
author = {Narayanan, Deepak and Harlap, Aaron and Phanishayee, Amar and Seshadri, Vivek and Devanur, Nikhil R. and Ganger, Gregory R. and Gibbons, Phillip B. and Zaharia, Matei},
title = {PipeDream: generalized pipeline parallelism for DNN training},
year = {2019},
isbn = {9781450368735},
publisher = {Association for Computing Machinery},
address = {New York, NY, USA},
url = {https://doi.org/10.1145/3341301.3359646},
doi = {10.1145/3341301.3359646},
booktitle = {Proceedings of the 27th ACM Symposium on Operating Systems Principles},
pages = {1–15},
numpages = {15},
location = {Huntsville, Ontario, Canada},
series = {SOSP '19}
}

@misc{glue_2019,
      title={GLUE: A Multi-Task Benchmark and Analysis Platform for Natural Language Understanding}, 
      author={Alex Wang and Amanpreet Singh and Julian Michael and Felix Hill and Omer Levy and Samuel R. Bowman},
      year={2019},
      eprint={1804.07461},
      archivePrefix={arXiv},
      primaryClass={cs.CL},
      url={https://arxiv.org/abs/1804.07461}, 
}

@misc{codet5_emnlp2021,
      title={CodeT5: Identifier-aware Unified Pre-trained Encoder-Decoder Models for Code Understanding and Generation}, 
      author={Yue Wang and Weishi Wang and Shafiq Joty and Steven C. H. Hoi},
      year={2021},
      eprint={2109.00859},
      archivePrefix={arXiv},
      primaryClass={cs.CL},
      url={https://arxiv.org/abs/2109.00859}, 
}

@misc{chatbot_2020,
      title={Towards a Human-like Open-Domain Chatbot}, 
      author={Daniel Adiwardana and Minh-Thang Luong and David R. So and Jamie Hall and Noah Fiedel and Romal Thoppilan and Zi Yang and Apoorv Kulshreshtha and Gaurav Nemade and Yifeng Lu and Quoc V. Le},
      year={2020},
      eprint={2001.09977},
      archivePrefix={arXiv},
      primaryClass={cs.CL},
      url={https://arxiv.org/abs/2001.09977}, 
}

@misc{questionanswering_emnlp2020,
      title={Dense Passage Retrieval for Open-Domain Question Answering}, 
      author={Vladimir Karpukhin and Barlas Oğuz and Sewon Min and Patrick Lewis and Ledell Wu and Sergey Edunov and Danqi Chen and Wen-tau Yih},
      year={2020},
      eprint={2004.04906},
      archivePrefix={arXiv},
      primaryClass={cs.CL},
      url={https://arxiv.org/abs/2004.04906}, 
}

@inproceedings{lsp-offload_aaai25,
  title={Practical Offloading for Fine-Tuning LLM on Commodity GPU via Learned Sparse Projectors},
  author={Chen, Siyuan and Wang, Zhuofeng and Guan, Zelong and Liu, Yudong and Gibbons, Phillip B},
  booktitle={Proceedings of the AAAI Conference on Artificial Intelligence},
  volume={39},
  number={22},
  pages={23614--23622},
  year={2025}
}

@misc{intrinsicdimensionalityexplainseffectiveness_2020,
      title={Intrinsic Dimensionality Explains the Effectiveness of Language Model Fine-Tuning}, 
      author={Armen Aghajanyan and Luke Zettlemoyer and Sonal Gupta},
      year={2020},
      eprint={2012.13255},
      archivePrefix={arXiv},
      primaryClass={cs.LG},
      url={https://arxiv.org/abs/2012.13255}, 
}

@inproceedings{ps_osdi2014,
author = {Li, Mu and Andersen, David G. and Park, Jun Woo and Smola, Alexander J. and Ahmed, Amr and Josifovski, Vanja and Long, James and Shekita, Eugene J. and Su, Bor-Yiing},
title = {Scaling distributed machine learning with the parameter server},
year = {2014},
isbn = {9781931971164},
publisher = {USENIX Association},
address = {USA},
booktitle = {Proceedings of the 11th USENIX Conference on Operating Systems Design and Implementation},
pages = {583–598},
numpages = {16},
location = {Broomfield, CO},
series = {OSDI'14}
}

@inproceedings{sgd_2010,
  title={Large-scale machine learning with stochastic gradient descent},
  author={Bottou, L{\'e}on},
  booktitle={Proceedings of COMPSTAT'2010: 19th International Conference on Computational StatisticsParis France, August 22-27, 2010 Keynote, Invited and Contributed Papers},
  pages={177--186},
  year={2010},
  organization={Springer}
}

@inproceedings{swapadvisor_asplos2020,
author = {Huang, Chien-Chin and Jin, Gu and Li, Jinyang},
title = {SwapAdvisor: Pushing Deep Learning Beyond the GPU Memory Limit via Smart Swapping},
year = {2020},
isbn = {9781450371025},
publisher = {Association for Computing Machinery},
address = {New York, NY, USA},
url = {https://doi.org/10.1145/3373376.3378530},
doi = {10.1145/3373376.3378530},
booktitle = {Proceedings of the Twenty-Fifth International Conference on Architectural Support for Programming Languages and Operating Systems},
pages = {1341–1355},
numpages = {15},
location = {Lausanne, Switzerland},
series = {ASPLOS '20}
}

@INPROCEEDINGS{sentinel_hpca2021,
  author={Ren, Jie and Luo, Jiaolin and Wu, Kai and Zhang, Minjia and Jeon, Hyeran and Li, Dong},
  booktitle={2021 IEEE International Symposium on High-Performance Computer Architecture (HPCA)}, 
  title={Sentinel: Efficient Tensor Migration and Allocation on Heterogeneous Memory Systems for Deep Learning}, 
  year={2021},
  volume={},
  number={},
  pages={598-611},
  doi={10.1109/HPCA51647.2021.00057}}

@article{aji2017sparse,
  title={Sparse communication for distributed gradient descent},
  author={Aji, Alham Fikri and Heafield, Kenneth},
  journal={arXiv preprint arXiv:1704.05021},
  year={2017}
}

@article{lin2017dgc,
  title={Deep gradient compression: Reducing the communication bandwidth for distributed training},
  author={Lin, Yujun and Han, Song and Mao, Huizi and Wang, Yu and Dally, William J},
  journal={arXiv preprint arXiv:1712.01887},
  year={2017}
}

@inproceedings{fedat_sc21,
  title={FedAT: A high-performance and communication-efficient federated learning system with asynchronous tiers},
  author={Chai, Zheng and Chen, Yujing and Anwar, Ali and Zhao, Liang and Cheng, Yue and Rangwala, Huzefa},
  booktitle={Proceedings of the international conference for high performance computing, networking, storage and analysis},
  pages={1--16},
  year={2021}
}

@inproceedings{stellaris_sc24,
  title={Stellaris: Staleness-Aware Distributed Reinforcement Learning with Serverless Computing},
  author={Yu, Hanfei and Wang, Hao and Tiwari, Devesh and Li, Jian and Park, Seung-Jong},
  booktitle={SC24: International Conference for High Performance Computing, Networking, Storage and Analysis},
  pages={1--17},
  year={2024},
  organization={IEEE}
}

@article{staleness_arxiv2018,
  title={Toward understanding the impact of staleness in distributed machine learning},
  author={Dai, Wei and Zhou, Yi and Dong, Nanqing and Zhang, Hao and Xing, Eric P},
  journal={arXiv preprint arXiv:1810.03264},
  year={2018}
}

@article{asynchronous_sgd_nips2015,
  title={Asynchronous parallel stochastic gradient for nonconvex optimization},
  author={Lian, Xiangru and Huang, Yijun and Li, Yuncheng and Liu, Ji},
  journal={Advances in neural information processing systems},
  volume={28},
  year={2015}
}

@inproceedings{convergence_PMLR2016,
  title={On convergence of model parallel proximal gradient algorithm for stale synchronous parallel system},
  author={Zhou, Yi and Yu, Yaoliang and Dai, Wei and Liang, Yingbin and Xing, Eric},
  booktitle={Artificial Intelligence and Statistics},
  pages={713--722},
  year={2016},
  organization={PMLR}
}

@inproceedings{petuum_kdd15,
  title={Petuum: A new platform for distributed machine learning on big data},
  author={Xing, Eric P and Ho, Qirong and Dai, Wei and Kim, Jin-Kyu and Wei, Jinliang and Lee, Seunghak and Zheng, Xun and Xie, Pengtao and Kumar, Abhimanu and Yu, Yaoliang},
  booktitle={Proceedings of the 21th ACM SIGKDD International Conference on Knowledge Discovery and Data Mining},
  pages={1335--1344},
  year={2015}
}

@inproceedings{deepspeed_kdd20,
  title={Deepspeed: System optimizations enable training deep learning models with over 100 billion parameters},
  author={Rasley, Jeff and Rajbhandari, Samyam and Ruwase, Olatunji and He, Yuxiong},
  booktitle={Proceedings of the 26th ACM SIGKDD international conference on knowledge discovery \& data mining},
  pages={3505--3506},
  year={2020}
}

@misc{risingcost_2025,
      title={The rising costs of training frontier AI models}, 
      author={Ben Cottier and Robi Rahman and Loredana Fattorini and Nestor Maslej and Tamay Besiroglu and David Owen},
      year={2025},
      eprint={2405.21015},
      archivePrefix={arXiv},
      primaryClass={cs.CY},
      url={https://arxiv.org/abs/2405.21015}, 
}

@misc{efficienttrainingsurvey_2023,
      title={On Efficient Training of Large-Scale Deep Learning Models: A Literature Review}, 
      author={Li Shen and Yan Sun and Zhiyuan Yu and Liang Ding and Xinmei Tian and Dacheng Tao},
      year={2023},
      eprint={2304.03589},
      archivePrefix={arXiv},
      primaryClass={cs.LG},
      url={https://arxiv.org/abs/2304.03589}, 
}

@inproceedings{
adamw_2019,
title={Decoupled Weight Decay Regularization},
author={Ilya Loshchilov and Frank Hutter},
booktitle={International Conference on Learning Representations},
year={2019},
url={https://openreview.net/forum?id=Bkg6RiCqY7},
}

@InProceedings{galore_2024,
  title = 	 {{G}a{L}ore: Memory-Efficient {LLM} Training by Gradient Low-Rank Projection},
  author =       {Zhao, Jiawei and Zhang, Zhenyu and Chen, Beidi and Wang, Zhangyang and Anandkumar, Anima and Tian, Yuandong},
  booktitle = 	 {Proceedings of the 41st International Conference on Machine Learning},
  pages = 	 {61121--61143},
  year = 	 {2024},
  editor = 	 {Salakhutdinov, Ruslan and Kolter, Zico and Heller, Katherine and Weller, Adrian and Oliver, Nuria and Scarlett, Jonathan and Berkenkamp, Felix},
  volume = 	 {235},
  series = 	 {Proceedings of Machine Learning Research},
  month = 	 {21--27 Jul},
  publisher =    {PMLR},
  url = 	 {https://proceedings.mlr.press/v235/zhao24s.html}
}

@misc{pytorch_nips2019,
      title={PyTorch: An Imperative Style, High-Performance Deep Learning Library}, 
      author={Adam Paszke and Sam Gross and Francisco Massa and Adam Lerer and James Bradbury and Gregory Chanan and Trevor Killeen and Zeming Lin and Natalia Gimelshein and Luca Antiga and Alban Desmaison and Andreas Köpf and Edward Yang and Zach DeVito and Martin Raison and Alykhan Tejani and Sasank Chilamkurthy and Benoit Steiner and Lu Fang and Junjie Bai and Soumith Chintala},
      year={2019},
      eprint={1912.01703},
      archivePrefix={arXiv},
      primaryClass={cs.LG},
      url={https://arxiv.org/abs/1912.01703}, 
}

@misc{deepspeed_flops_profiler_2025,
  title        = {DeepSpeed Flops Profiler},
  author       = {{DeepSpeed Team}},
  year         = {2025},
  howpublished = {https://www.deepspeed.ai/tutorials/flops-profiler/} 
}

@misc{runpod-pricing,
  author       = {RunPod},
  title        = {RunPod Pricing},
  howpublished = {\url{https://www.runpod.io/pricing}},
}

@article{qlora_nips2023,
  title={Qlora: Efficient finetuning of quantized llms},
  author={Dettmers, Tim and Pagnoni, Artidoro and Holtzman, Ari and Zettlemoyer, Luke},
  journal={Advances in neural information processing systems},
  volume={36},
  pages={10088--10115},
  year={2023}
}

@article{lora_iclr2022,
  title={Lora: Low-rank adaptation of large language models.},
  author={Hu, Edward J and Shen, Yelong and Wallis, Phillip and Allen-Zhu, Zeyuan and Li, Yuanzhi and Wang, Shean and Wang, Lu and Chen, Weizhu and others},
  journal={ICLR},
  volume={1},
  number={2},
  pages={3},
  year={2022}
}

@article{sparseadapter_arxiv2022,
  title={Sparseadapter: An easy approach for improving the parameter-efficiency of adapters},
  author={He, Shwai and Ding, Liang and Dong, Daize and Zhang, Miao and Tao, Dacheng},
  journal={arXiv preprint arXiv:2210.04284},
  year={2022}
}

@article{prefix_tuning_arxiv2021,
  title={Prefix-tuning: Optimizing continuous prompts for generation},
  author={Li, Xiang Lisa and Liang, Percy},
  journal={arXiv preprint arXiv:2101.00190},
  year={2021}
}

@inproceedings{asynchronous_sgd_iclr2018,
  title={Asynchronous decentralized parallel stochastic gradient descent},
  author={Lian, Xiangru and Zhang, Wei and Zhang, Ce and Liu, Ji},
  booktitle={International conference on machine learning},
  pages={3043--3052},
  year={2018},
  organization={PMLR}
}

@article{elastic_averaging_sgd_nips2015,
  title={Deep learning with elastic averaging SGD},
  author={Zhang, Sixin and Choromanska, Anna E and LeCun, Yann},
  journal={Advances in neural information processing systems},
  volume={28},
  year={2015}
}

@inproceedings{g10_micro2023,
  title={G10: Enabling an efficient unified gpu memory and storage architecture with smart tensor migrations},
  author={Zhang, Haoyang and Zhou, Yirui and Xue, Yuqi and Liu, Yiqi and Huang, Jian},
  booktitle={Proceedings of the 56th Annual IEEE/ACM International Symposium on Microarchitecture},
  pages={395--410},
  year={2023}
}

@misc{superoffload_asplos2026,
      title={SuperOffload: Unleashing the Power of Large-Scale LLM Training on Superchips}, 
      author={Xinyu Lian and Masahiro Tanaka and Olatunji Ruwase and Minjia Zhang},
      year={2025},
      eprint={2509.21271},
      archivePrefix={arXiv},
      primaryClass={cs.LG},
      url={https://arxiv.org/abs/2509.21271}, 
}

@inproceedings{revisiting_acl2022,
  title={Revisiting Parameter-Efficient Tuning: Are We Really There Yet?},
  author={Chen, Guanzheng and Liu, Fangyu and Meng, Zaiqiao and Liang, Shangsong},
  booktitle={Proceedings of the 2022 Conference on Empirical Methods in Natural Language Processing},
  editor={Goldberg, Yoav and Kozareva, Zornitsa and Zhang, Yue},
  pages={2612--2626},
  address={Abu Dhabi, United Arab Emirates},
  month={December},
  year={2022},
  publisher={Association for Computational Linguistics},
  doi={10.18653/v1/2022.emnlp-main.168},
  url={https://aclanthology.org/2022.emnlp-main.168/}
}

@inproceedings{parameter_acl2023,
  title={Parameter-Efficient Fine-Tuning without Introducing New Latency},
  author={Liao, Baohao and Meng, Yan and Monz, Christof},
  booktitle={Proceedings of the 61st Annual Meeting of the Association for Computational Linguistics (Volume 1: Long Papers)},
  editor={Rogers, Anna and Boyd-Graber, Jordan and Okazaki, Naoaki},
  pages={4242--4260},
  address={Toronto, Canada},
  month={July},
  year={2023},
  publisher={Association for Computational Linguistics},
  doi={10.18653/v1/2023.acl-long.233},
  url={https://aclanthology.org/2023.acl-long.233/}
}

@article{parameter_2025arxiv,
  title={Parameter-Efficient Continual Fine-Tuning: A Survey},
  author={Coleman, Eric Nuertey and Quarantiello, Luigi and Liu, Ziyue and Yang, Qinwen and Mukherjee, Samrat and Hurtado, Julio and Lomonaco, Vincenzo},
  journal={arXiv preprint arXiv:2504.13822},
  year={2025},
  month={April},
  note={Version 2, August 2025},
  url={https://arxiv.org/abs/2504.13822}
}

@article{dai2024discovering,
  title={Discovering Long-Term Effects on Parameter Efficient Fine-tuning},
  author={Dai, Gaole and Fan, Chun-Kai and Tang, Yiming and Zhang, Qizhe and Gan, Yulu and Zhang, Zhi and Zeng, Chengqing and Zhang, Shanghang and Huang, Tiejun},
  journal={arXiv preprint arXiv:2409.06706},
  year={2024},
  month={September},
  note={Submitted to ICLR 2025},
  url={https://arxiv.org/abs/2409.06706}
}

@misc{patel2025guide,
  title={Guide to fine-tuning LLMs using PEFT and LoRA techniques},
  author={Patel, Pranav},
  year={2025},
  howpublished={Mercity Blog},
  note={\url{https://www.mercity.ai/blog-post/fine-tuning-llms-using-peft-and-lora}},
  url={https://www.mercity.ai/blog-post/fine-tuning-llms-using-peft-and-lora}
}

@inproceedings{maskllm_nips2024,
author = {Fang, Gongfan and Yin, Hongxu and Muralidharan, Saurav and Heinrich, Greg and Pool, Jeff and Kautz, Jan and Molchanov, Pavlo and Wang, Xinchao},
title = {MaskLLM: learnable semi-structured sparsity for large language models},
year = {2024},
isbn = {9798331314385},
publisher = {Curran Associates Inc.},
address = {Red Hook, NY, USA},
booktitle = {Proceedings of the 38th International Conference on Neural Information Processing Systems},
articleno = {248},
numpages = {23},
location = {Vancouver, BC, Canada},
series = {NIPS '24}
}

@misc{huang2025castcontinuousdifferentiablesemistructured,
      title={CAST: Continuous and Differentiable Semi-Structured Sparsity-Aware Training for Large Language Models}, 
      author={Weiyu Huang and Yuezhou Hu and Jun Zhu and Jianfei Chen},
      year={2025},
      eprint={2509.25996},
      archivePrefix={arXiv},
      primaryClass={cs.LG},
      url={https://arxiv.org/abs/2509.25996}, 
}

@inproceedings{alistarh2018convergence,
  title={The convergence of sparsified gradient methods},
  author={Alistarh, Dan and Hoefler, Torsten and Johansson, Mikael and Konstantinov, Nikola and Khirirat, Sarit and Renggli, Cedric},
  booktitle={Advances in Neural Information Processing Systems},
  volume={31},
  year={2018}
}

@inproceedings{stich2018sparsified,
  title={Sparsified SGD with memory},
  author={Stich, Sebastian U and Cordonnier, Jean-Baptiste and Jaggi, Martin},
  booktitle={Advances in Neural Information Processing Systems},
  volume={31},
  year={2018}
}

@article{m-lora_vldb25,
author = {Ye, Zhengmao and Li, Dengchun and Hu, Zetao and Lan, Tingfeng and Sha, Jian and Zhang, Shicong and Duan, Lei and Zuo, Jie and Lu, Hui and Zhou, Yuanchun and Tang, Mingjie},
title = {mLoRA: Fine-Tuning LoRA Adapters via Highly-Efficient Pipeline Parallelism in Multiple GPUs},
year = {2025},
issue_date = {February 2025},
publisher = {VLDB Endowment},
volume = {18},
number = {6},
issn = {2150-8097},
url = {https://doi.org/10.14778/3725688.3725718},
doi = {10.14778/3725688.3725718},
journal = {Proc. VLDB Endow.},
month = feb,
pages = {1948–1961},
numpages = {14}
}

@article{dlrover-rm_vldb25,
author = {Wang, Qinlong and Lan, Tingfeng and Tang, Yinghao and Sang, Bo and Huang, Ziling and Du, Yiheng and Zhang, Haitao and Sha, Jian and Lu, Hui and Zhou, Yuanchun and Zhang, Ke and Tang, Mingjie},
title = {DLRover-RM: Resource Optimization for Deep Recommendation Models Training in the Cloud},
year = {2024},
issue_date = {August 2024},
publisher = {VLDB Endowment},
volume = {17},
number = {12},
issn = {2150-8097},
url = {https://doi.org/10.14778/3685800.3685832},
doi = {10.14778/3685800.3685832},
journal = {Proc. VLDB Endow.},
month = aug,
pages = {4130–4144},
numpages = {15}
}
